\documentclass[usenatbib,twocolumn]{mn2e}
\usepackage[]{graphicx}

\def\keV{\thinspace \rm{keV}}

\def\cm{{\rm\thinspace cm}}

\def\erg{{\rm\thinspace erg}}
\def\s{{\rm\thinspace s}}

\def\ergpcmsq{{\rm \erg\cm^{-2}}}

\def\swift{{\it Swift}}

\title[The early time emission of GRB~080310]
{The origin of the early time optical emission of Swift
 GRB~080310\thanks{Based on observations made also with ESO Telescopes at the
 La Silla and Paranal Observatory under programme IDs 080.D-0250 and
 080.D-0791}}

\author[O.M. Littlejohns et al.]
{O.M. Littlejohns$^{1}$,\thanks{E-mail: oml2@star.le.ac.uk (OML)}
R. Willingale$^{1}$,
P.T. O'Brien$^{1}$,
A.P. Beardmore$^{1}$,
S. Covino$^{2}$,
\and
D.A. Perley$^{3}$,
N.R. Tanvir$^{1}$,
E. Rol$^{4}$,
F. Yuan$^{5}$,
C. Akerlof$^{6}$,
P. D'Avanzo$^{2}$,
\and
D.F. Bersier$^{7}$,
A.J. Castro-Tirado$^{8}$,
P. Christian$^{3}$,
B.E. Cobb$^{9}$,
P.A. Evans$^{1}$,
\and
A.V. Filippenko$^{3}$,
H. Flewelling$^{10}$,
D. Fugazza$^{2}$,
E.A. Hoversten$^{11}$,
A.P. Kamble$^{4}$ 
\and
S. Kobayashi$^{7}$,
W. Li$^{3}$,
A.N. Morgan$^{3}$,
C.G. Mundell$^{7}$,
K. Page$^{1}$,
E. Palazzi$^{12}$,
\and
R.M. Quimby$^{13}$,
S. Schulze$^{14}$,
I.A. Steele$^{7}$,
A. de Ugarte Postigo$^{15}$
\\
$^{1}$ Department of Physics and Astronomy, University of Leicester,
LE1~7RH, UK\\
$^{2}$ INAF/Osservatorio Astronomico di Brera, via Emilio Bianchi 46, 23807 Merate (LC), Italy\\
$^{3}$ Department of Astronomy, University of California, Berkeley, CA 94720-3411, USA\\
$^{4}$ Astronomical Institute ``Anton Pannekoek'', P.O. Box 94248, NL-1090 SJ Amsterdam, The Netherlands\\
$^{5}$ Research School of Astronomy and Astrophysics, The Australian National University, Cotter Road, Weston Creek, ACT 2611, Australia\\
$^{6}$ Physics Department, University of Michigan, Ann Arbor, MI 48109, USA\\
$^{7}$ Astrophysics Research Institute, Liverpool John Moores University, Twelve Quays House, Egerton Wharf, Birkenhead, CH41~1LD, UK\\
$^{8}$ Instituto de Astrofisica de Andaluca (IAA-CSIC), P.O. Box 03004, E-18008, Granada, Spain\\
$^{9}$ Department of Physics, The George Washington University, Corcoran 105, 725 21st St, NW, Washington, DC 20052, USA\\
$^{10}$ Institute for Astronomy, University of Hawaii at Manoa, Honolulu, HI 96822, USA\\
$^{11}$ Department of Astronomy \& Astrophysics, The Pennsylvania State
  University, 525 Davey Laboratory, University Park, PA 16802, USA\\
$^{12}$ INAF - IASF di Bologna, via Gobetti 101, 40129 Bologna, Italy\\
$^{13}$ Cahill Center for Astrophysics 249-17, California Institute of Technology, Pasadena, CA 91125, USA\\
$^{14}$ Centre for Astrophysics and Cosmology, Science Institute, University of Iceland, Dunhagi 5, 107 Reykjavk, Iceland\\
$^{15}$ Dark Cosmology Centre, Niels Bohr Institute, University of Copenhagen, Juliane Maries Vej 30, 2100 Copenhagen, Denmark\\
}
\begin{document}
\date{2011 Dec 31}
%\date{Accepted 2005 June 15. Received 1988 December 14; in original form 1988 October 11}

\pagerange{\pageref{firstpage}--\pageref{lastpage}} \pubyear{2011}

\maketitle

\label{firstpage}

\begin{abstract}
We present broadband multi-wavelength observations of GRB~080310 at redshift
 $z$ $=$ $2.43$. This burst was bright and long-lived, and unusual in having
 extensive optical and near IR follow-up during the prompt phase. Using these
 data we attempt to simultaneously model the $\gamma$-ray, X-ray, optical and
 IR emission using a series of prompt pulses and an afterglow component.
 Initial attempts to extrapolate the high energy model directly to lower
 energies for each pulse reveal that a spectral break is required
 between the optical regime and 0.3 keV to avoid over predicting the optical
 flux. We demonstrate that afterglow emission alone is insufficient to
 describe all morphology seen in the optical and IR data. Allowing the
 prompt component to dominate the early-time optical and IR and permitting
 each pulse to have an independent low energy spectral indices we produce an
 alternative scenario which better describes the optical light curve. This,
 however, does not describe the spectral shape of GRB~080310 at early times.
 The fit statistics for the prompt and afterglow dominated models are nearly
 identical making it difficult to favour either. However one enduring result
 is that both models require a low energy spectral index consistent with self
 absorption for at least some of the pulses identified in the high energy
 emission model.
\end{abstract}


\section{Introduction}

Over the last few years a combination of fast-response ground-based telescopes
 triggered by the availability of rapid, accurate localisations have started
 to provide the data required to answer the question of what is causing the
 early, bright X-ray and optical emission from gamma-ray bursts (GRBs). The
 most accurate prompt X-ray locations come from the \swift\ satellite
 \citep{2004ApJ...611.1005G}. These are supplemented by either on-board or
 ground detections of the ultraviolet (UV), optical or infra-red (IR)
 counterpart.\par
In the popular relativistic fireball model for GRBs, the early, usually highly
 variable emission is understood to be due to internal shocks
 \citep{1997ApJ...485..270S} or magnetic dissipation within the jet, and the
 so-called $external$ emission is produced by the interaction of the jet with
 the surrounding medium. The latter emission is usually described using the
 fireball model \citep{1992MNRAS.258P..41R}, which has successfully been
 applied to describe the behaviour of GRBs half a day or so after the trigger,
 but has difficulties explaining the complex behaviour seen in the first few
 hours, a period now routinely accessed by \swift\ and other rapid-response
 facilities. Ideally multi-wavelength observations should be obtained while
 the burst is happening so as to try to disentangle the relative contribution
 from the internal and external components.\par
\citet{2009MNRAS.397.1177E} present a uniformly analysed comprehensive sample
 of 317 \swift\ GRBs spanning from December 2004 to July 2008, in which their
 morphologies are compared to the proposed canonical X-ray light curve
 (\citealt{2006ApJ...642..389N}, \citealt{2006ApJ...642..354Z} \&
 \citealt{2006MNRAS.366.1357P}). Such canonical light curves consider the
 X-ray emission to consist of a series of power-laws, where one important
 phase is the rapid decay phase which has been explained as being the
 smooth continuation of the prompt emission (\citealt{2006ApJ...647.1213O},
 \citealt{2005Natur.436..985T} \& \citealt{2005ApJ...635L.133B}). From the
 sample of \citet{2009MNRAS.397.1177E} it is clear that the X-ray light curves
 of GRBs vary from burst to burst. Some show strong flaring, one example being
 GRB~061121 \citep{2007ApJ...663.1125P}, however, some bursts show remarkably
 simple and smooth decay (GRB~061007; \citealt{2007MNRAS.380.1041S}). Similar
 findings are also reported in \citet{2009ApJ...698...43R}. Rapid behaviour,
 such as flaring, at high energies is often attributed to central-engine
 behaviour \citep{2010MNRAS.406.2149M} but how this relates to the optical
 emission remains somewhat of a mystery. The available datasets reveal a
 confusing picture. In some cases the early optical data seem to trace the
 X-ray and $\gamma$-ray light curves, such as GRB~041219A
 (\citealt{2005Natur.435..178V} \& \citealt{2005Natur.435..181B}),
 suggesting that optical flaring may be of internal origin. In other GRBs the
 optical behaviour seems entirely unrelated to the the high-energy emission
 (GRB~990123; \citealt{1999Natur.398..400A} \& GRB~060607A;
 \citealt{2009ApJ...693.1417N} \& \citealt{2007A&A...469L..13M}), and instead
 seems to follow the behaviour of the external afterglow.\par
To make progress requires continued efforts to observe GRBs over as wide a
 wavelength range as possible from as early as possible. This is only really
 viable for bright, long-lasting GRBs which are well-placed for rapid
 follow-up. Here we present prompt, multi-wavelength data from the GRB~080310,
 which begin at a time soon after the trigger that is not often accessed by
 ground-based facilities. Following the trigger, GRB~080310 was
 detected on-board \swift\ by both the X-Ray Telescope (XRT)
 \citep{2005SSRv..120..165B} and UV/Optical Telescope (UVOT)
 \citep{2005SSRv..120...95R} and also observed in the optical and IR by
 several ground-based telescopes. These rare data present us with an
 opportunity to discriminate between whether the early time lower-energy
 light curve of a GRB is driven by internal or external emission, during which
 time the high-energy emission is presumed to be totally internally
 dominated.\par
In the following section we discuss the observations, then in \S 3 we present
 attempts to fit both an internal shock model \citep{2009MNRAS.399.1328G} and
 an afterglow component \citep{2007ApJ...662.1093W} to the X-ray and
 $\gamma$-ray data, and describe the necessary modification required to
 simultaneously fit the early optical emission in two scenarios, where either
 the prompt or afterglow components dominate this early flux. In \S 4 we
 briefly discuss the relative merits of using each component, before finally
 in \S 5 we conclude which of the two alternatives is a better fit to the
 observed data and consider the implications of the model on the physics
 governing the emission from GRB~080310.\par

\section{Observations}

\begin{figure}
\begin{center}
\includegraphics[width=8cm,angle=0]{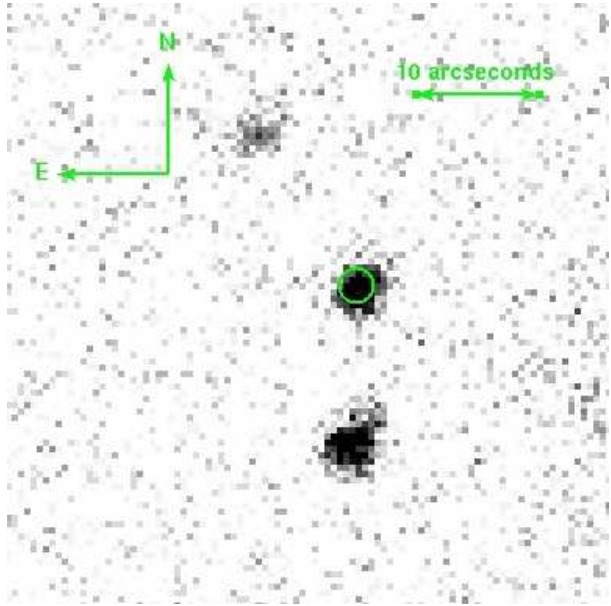}
\end{center}
\caption{UVOT $v$-band image from $T_0+79\s$ to $T_0+1539\s$. The enhanced
 \swift\ XRT position is shown \citep{2008GCN..7394....1O}, with an error
 circle of radius $1\farcs4$, corresponding to the 90\% confidence limit,
 containing the optical afterglow candidate at an average $v$-band magnitude
 of 17.49.}
\label{v_img}
\end{figure}

\begin{figure*}
\begin{center}
\includegraphics[height=10.0cm]{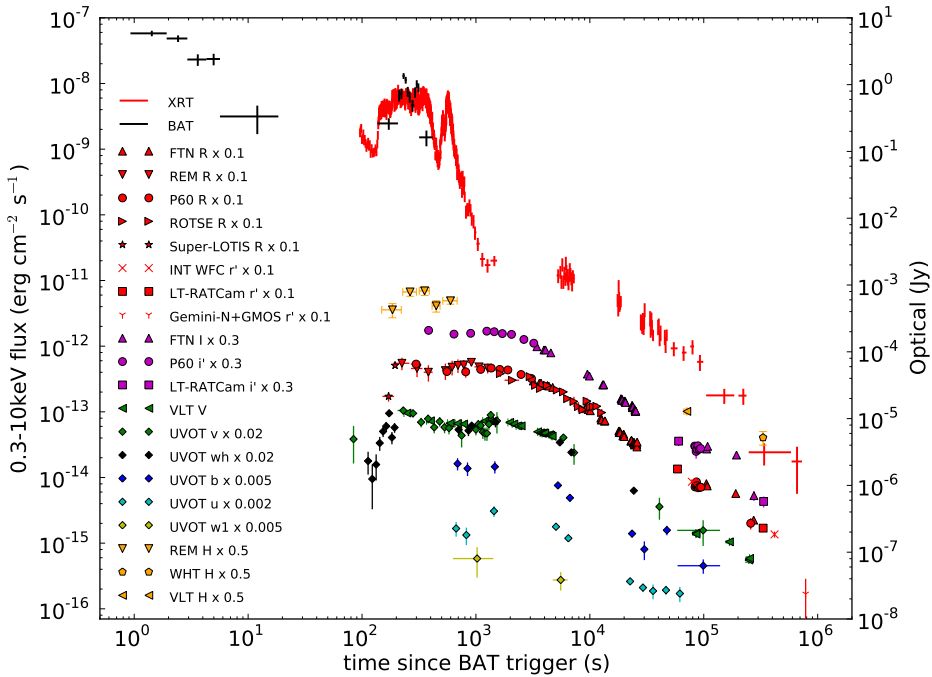}
\end{center}
\caption{$\gamma$-ray to near IR light curves for GRB~080310, showing the
 energy bands used to observe the GRB and the instruments used to take the
 measurements. Data have been scaled from their actual values for viewing
 purposes.}
\label{all_lc}
\end{figure*}

On 2008 March 10 the \swift\ BAT triggered and located GRB~080310 (trigger
 number 305288) on board at 08:37:58 UT \citep{2008GCN..7382....1C}. \swift\
 slewed immediately which enabled the narrow field instruments to begin
 observing the burst $89\s$ after the trigger. The burst was detected by the
 XRT and UVOT (\textit{white} filter), with the latter providing the best
 \swift\ position of ${\rm RA(J2000)} = 14^{h}40^{m}13\fs80$,
 ${\rm Dec(J2000)} = -00\degr 10\arcmin29\farcs6$ with a 1$\sigma$ error
 radius of $0\farcs6$. Figure \ref{v_img} shows a UVOT $v$-band image from the
 early time data. The \swift\ light curves obtained in multiple bands
 from each of the three on board instruments are presented in Figure
 \ref{swift_lcs} and most of the available datasets from an extensive number
 of facilities are shown in Figure \ref{all_lc}. Data from the PAIRITEL 
\citep{2008GCN..7406....1P} and KAIT  instruments are shown separately in
 Figure \ref{KP}.\par
%\begin{figure}
%\begin{center}
%\includegraphics[width=8cm,angle=0]{xrt_image.ps}
%\end{center}
%\caption{XRT image over all nine XRT observational segments, giving a total
% exposure time of 106.9 ks. The circular region is the \swift\
% enhanced XRT position \citep{2008GCN..7394....1O}, and with an error radius
% of $1\farcs4$ (90\% confidence).}
%\label{xrt_img}
%\end{figure}
In addition to \swift\ observations, and those instruments already mentioned,
 GRB~080310 was also observed on the ground with numerous optical and NIR
 facilities, including REM \citep{2008GCN..7385....1C}, VLT
 \citep{2008GCN..7393....1C} and the Faulkes Telescope North. These
 observations are shown alongside the BAT and XRT light curves in Figure
 \ref{all_lc}.\par
The Kast dual spectrometer at the Lick Observatory, California, obtained the
 first redshift estimation for this burst of $z$ = 2.4266
 \citep{2008GCN..7388....1P} using strong absorption features from Silicon,
 Carbon and Aluminium. This was later corroborated by the VLT-UVES instrument
 \citep{2008GCN..7391....1V} and the Keck-DEIMOS spectrometer
 \citep{2008GCN..7397....1P}.\par
The following subsections describe the observations in more detail. The
 \swift\ data analysis was performed using release 2.8 of the \swift\ software
 tools. Parameter uncertainties are estimated at the 90\% confidence level. We
 note that the optical datasets have been reduced using different methods, and
 fully investigated the effects of cross calibration errors, ensuring that our
 later analysis remained insensitive to their effects.\par

\subsection{BAT}

\begin{figure}
\begin{center}
\includegraphics[width=9.0cm]{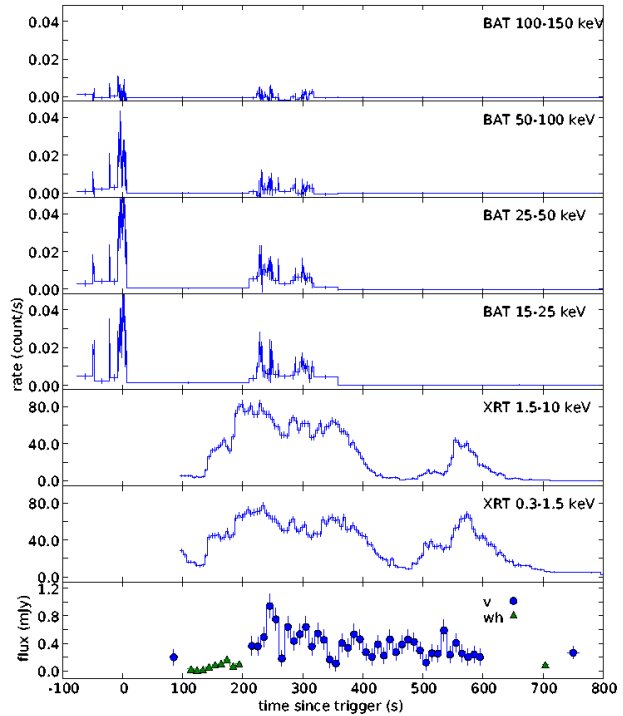}
\end{center}
\caption{\swift\ prompt light curves from GRB~080310 in different energy bands.
 From top to bottom, BAT $100-150 \keV$, BAT $50-100 \keV$, BAT $25-50 \keV$,
 BAT $15-25 \keV$, XRT $1.5-10.0 \keV$, XRT $0.3-1.5 \keV$ and UVOT $v$ and
 \textit{white} bands. Strictly, the count rates for BAT are per fully
 illuminated detector.}
\label{swift_lcs}
\end{figure}

The BAT data were processed using the standard {\sl batgrbproduct} script. The
 top four panels of Figure \ref{swift_lcs} show the BAT light curves displayed
 in the standard energy bands of $15-25\keV$, $25-50\keV$, $50-100\keV$ and
 $100-150\keV$, plotted with respect to the BAT trigger time ($T_0$). The
 binning is such that each bin satisfies a minimum signal-to-noise ratio of 5
 and a minimum time bin size of $1\s$.\par
The $\gamma$-ray light curve shows many peaks with the first at $T_0 - 60\s$.
 The brightest peak extends from $T_0 - 12\s$ to $T_0 + 7\s$. This is followed
 by a period of no detectable emission before a weaker, broad series of peaks
 is seen from $T_0 +180\s$ to $T_0 +360\s$ \citep{2008GCN..7402....1T}. The
 latter peak is consistent with the first strong flare seen in the XRT (see
 below). The BAT emission is strongest in the lower energy bands, below
 $\sim 100\keV$. The $T_{90}$ is estimated to be $365 \pm 20 \s$ (where the
 error includes systematics).\par
The total spectrum from $T_0 - 71.76\s$ to $T_0 + 318.75\s$ is well fit by
 a power-law of photon index $2.32 \pm 0.16$, with a total fluence of
 $2.3\pm0.2 \times 10^{-6} \ergpcmsq$ over the $15-150 \keV$ band. The
 fluence ratio $S(25-50\keV)/S(50-100\keV)$ is $1.27\pm0.17$ which puts
 GRB~080310 on the border of the X-ray-rich gamma-ray bursts and X-ray flash
 (XRF) according to the definition of \citet{2008ApJ...679..570S}.\par

\subsection{XRT}

The \swift\ XRT started observations of GRB~080310 $89 \s$ after the trigger,
 with windowed timing data ranging from $T_0+95 \s$ to $T_0+799 \s$
 and photon counting mode data after this \citep{2008GCN..7399....1B}. As can
 be seen in Figure \ref{swift_lcs} the soft band XRT light curve initial fades
 until $T_0+130 \s$, before the burst rebrightens. This flaring activity
 continues until $T_0+420 \s$ before the count rate drops again briefly. A
 further flaring event is seen between $T_0+500 \s$ and $T_0+620 \s$, which is
 approximately half as bright as the first. There is significant spectral
 evolution during the flaring events, where the $S(1.5-10 \keV)/S(0.3-1.5
 \keV)$ ratio shows a hardening in the spectrum at $T_0+135 \s$ to $T_0+200
 \s$ and $T_0+500 \s$ to $T_0+565 \s$ before, in both instances, there is a
 softening as the flaring behaviour declines.\par

\subsection{UVOT}

UVOT observations began with a 100 second finding chart exposure taken at
 $T_0+99 \s$ \citep{2008GCN..7398....1H}. This finding chart exposure was the
 sum of 10 individual 10 second exposures, with GRB~080310 being detected in
 all but the first exposure. At the time of GRB~080310
 UVOT observations during the first orbit were taken in the event mode, which
 allows for higher timing resolution, while during subsequent orbits
 observations were taken in imaging mode. UVOT photometry was done using the
 publicly available \textit{FTOOLS} data reduction suite, and is in the
 UVOT photometric system described in \citet{2008MNRAS.383..627P}. During the
 first 1,000 seconds the light curve is complex as shown in Figure
 \ref{swift_lcs}, after which the the burst can be seen to fade in all the
 observed optical bands (Figure \ref{all_lc}).\par
There are available data for the UVOT $white$, $v$, $b$, $u$ and $uvw1$.
 As this burst has a measured redshift of
 $z~=~2.43$ it is important to consider absorption from the intervening medium
 through which the emission must travel. Correction factors were calculated for
 Lyman absorption using the $GRBz$ code described in
 \citet{2008A&A...490.1047C}. $GRBz$ uses the model presented by
 \citet{1995ApJ...441...18M} to calculate the absorption from neutral
 Hydrogen in the intergalactic medium. Having found the correction factor for
 all the optical and near infrared bands presented in this work, we noted that
 it was only necessary to correct the data in the $b$, $u$ and $uvw1$ bands.
 Whilst the correction to both the $u$ and $b$ bands was not large, we found
 that the $uvw1$ data required a significant correction, within which there
 was a large uncertainty. Given that there were only two data points from UVOT
 in this band, we decided to remove them from any further analysis.\par

\subsection{Faulkes Telescope North}

Observations with the Faulkes Telescope North (FTN) started at 2008-03-10
 09:31:07.3 (UT), 3.188 ks after the trigger. Data were reduced in a standard
 fashion using {\sc iraf} \citep{1986SPIE..627..733T}. Calibration was
 performed using the SDSS data for the region \citep{2007ApJS..172..634A}. For
 the $I$ and $R$ filters, the SDSS photometry was converted to the
 Johnson-Cousins system.\footnote{See Lupton (2005) at
 \\ http://www.sdss.org/dr6/algorithms/sdssUBVRITransform.html} Photometry
 was then performed using an aperture matched to the average seeing of the
 (combined) frames. For the conversion from magnitude to flux, the data were
 first corrected for Galactic extinction using the COBE-DIRBE extinction maps
 from \citet{1998ApJ...500..525S}, and then converted using flux zero points
 from \citet{1995PASP..107..945F} for optical and \citet{2005PASP..117..421T}
 for infrared. AB magnitudes were converted following
 \citet{1983ApJ...266..713O}.\par

\subsection{ROTSE}

ROTSE-IIIb, located at McDonald Observatory, Texas, responded to GRB~080310
 and began imaging 5.7 seconds after the GCN notice time \citep{gcn7411}.
 Observations were carried out under fluctuating weather conditions. The
 optical transient (OT) was detected between 25 minutes and 3.5 hours after
 the trigger. To improve detection signal to noise ratio, sets of 4 to 11
 images are co-added and exposures badly affected by weather are excluded. The
 OT is slightly blended with two nearby stars in the ROTSE images. We
 therefore subtract the scaled point spread functions (PSFs) of these two
 nearby stars and then apply PSF-matching photometry on the OT using our
 custom RPHOT package \citep{qryaa06}. The analysis is further complicated by
 large seeing variation, particularly towards the end of the observation. The
 structure seen in the light curve during this time is likely not significant.
 The ROTSE-III unfiltered magnitudes are calibrated to SDSS $r$ using standard
 stars in the pre-burst SDSS observations \citep{gcn7396}.\par

\subsection{REM}

Early time optical and NIR data were collected using the 60-cm robotic
 telescope REM (\citealt{2001AN....322..275Z}; \citealt{2004SPIE.5492.1613C})
 located at the European Southern Observatory (ESO) La Silla observatory
 (Chile). The telescope simultaneously feeds, by means of a dichroic, the two
 focal instruments: REMIR \citep{2004SPIE.5492.1602C} a NIR camera, operating
 in the range 1.0 to 2.3 $\mu$m ($z'$, $J$, $H$ and $K$) and ROSS (REM Optical
 Slitless Spectrograph; \citealt{2004SPIE.5492..689T}) an optical imager with
 spectroscopic (slitless) and photometric capabilities ($V$, $R$, $I$). Both
 cameras have a field of view of 10 $\times$ 10 arcmin$^2$.\par
REM reacted automatically after receiving the \textit{Swift} alert for
 GRB~080310, and began observing about 150 seconds after the GRB trigger time
\citep{2008GCN..7385....1C}.\par
Optical and NIR data were reduced following standard procedures. In
 particular, each single NIR observation with REMIR was performed with a
 dithering sequence of five images shifted by a few arcseconds. These images
 are automatically elaborated using the proprietary routine AQuA
 \citep{2004SPIE.5496..729T}. The script aligns the images and co-adds all
 the frames to obtain one average image for each sequence. Astrometry was
 performed using USNO-B1\footnote{http://tdc-www.harvard.edu/catalogs/ub1.html}
 and 2MASS\footnote{http://pegasus.phast.umass.edu/} catalogue reference
 stars.\par
Photometry was derived by a combination of the SExtractor package
 \citep{1996A&AS..117..393B} and the photometric tools provided by the
 {\sc gaia}\footnote{http://star-www.dur.ac.uk/~pdraper/gaia/gaia.html}
 package. The photometric calibration for the NIR was accomplished by applying
 average magnitude shifts computed using 2MASS isolated and non-saturated
 stars. The optical data were calibrated using instrumental zero points and
 checked with observations of standard stars in the field provided by
 \citet{2008GCN..7631....1H}.\par

\subsection{VLT}

VLT FORS1 $V$ and $R$ observations for GRB~080310 were automatically activated
 with the RRM mode\footnote{http://www.eso.org/sci/observing/phase2/SMSpecial
/RRMObservation.html} allowing the telescope to react promptly to any alert.
 The field was acquired and the observations began less than seven minutes
 after the GRB trigger. Later VLT observations were obtained with ISAAC at
 about one day after the burst with the $J$, $H$ and $K$ filters. In addition
 linear polarimetry observations were carried out with FORS1 with the $V$
 filter at approximately one, two and three days after the trigger.\par
Optical and NIR data were reduced following standard procedures with the
 tools of the ESO-eclipse package \citep{1997Msngr..87...19D}. Polarimetric
 data were reduced again following standard procedures as discussed in
 \citet{1999A&A...348L...1C,2002A&A...392..865C,2003A&A...400L...9C}.
 Photometry was performed by means of the tools provided by the {\sc gaia}
 package and with PSF photometry with the ESO-midas
\footnote{http://www.eso.org/sci/software/esomidas/} DAOPHOT context
 \citep{1987PASP...99..191S}.\par
The photometric calibration for the NIR was accomplished by applying average
 magnitude shifts computed using 2MASS isolated and non-saturated stars. The
 optical data were calibrated using instrumental zero points and checked with
 observations of standard stars in the field provided by
 \citet{2008GCN..7631....1H}. Linear polarimetry position angle was corrected
 by means of observations of polarimetric standard stars in the NGC 2024
 region.\par

\subsection{WHT}

Late imaging was obtained with the 4.2m William Herschel Telescope (WHT), at
 Roque de los Muchachos Observatory (La Palma, Spain) using the Long-slit
 Intermediate Resolution Infrared Spectrograph (LIRIS) in its imaging mode.
 Observations consisted of $36\times25$ second exposures in $H$-band, obtained
 on the 14th March 2008 from 04:36:28 to 04:53:41 UT. The data were reduced
 following standard procedures in {\sc iraf}. For the photometric calibration,
 we used stars from the 2MASS catalogue as reference.\par

\subsection{PAIRITEL}

PAIRITEL \citep{2006ASPC..351..751B} responded to GRB 080310 and began
 taking data at  09:04:58 (UT) in the $J$, $H$, and $K$ filters simultaneously
 \citep{2008GCN..7406....1P}.  The afterglow \citep{2008GCN..7381....1C} was
 well-detected in all three filters. \citet{2008GCN..7406....1P} also report
 on an SED constructed using data from PAIRITEL, KAIT and UVOT
 \citep{2008GCN..7398....1H}, allowing a joint fit to be made and the
 estimation of a small amount ($A_{V}$ = 0.10 $\pm$0.05) of SMC-like
 host-galaxy extinction.\par

\subsection{KAIT}

\begin{figure}
\begin{center}
\includegraphics[width=8cm,height=8cm,angle=270]{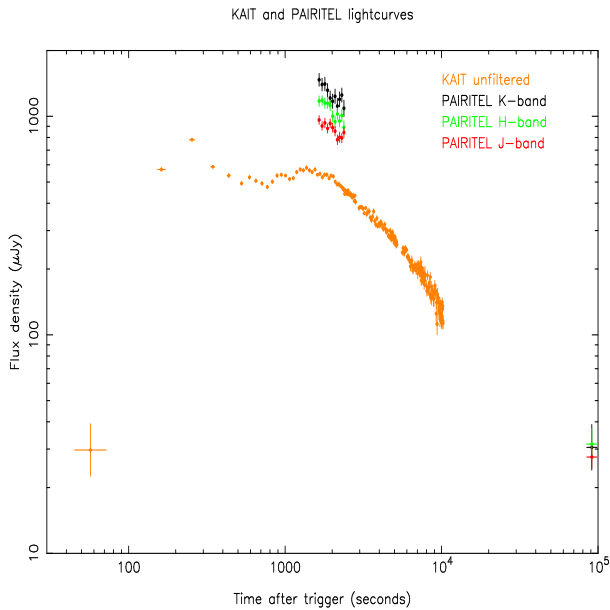}
\end{center}
\caption{PAIRITEL and KAIT light curves for GRB~080310. The KAIT unfiltered
 data are calibrated to the $R$-band. All data have been corrected for dust
 extinction.}
\label{KP}
\end{figure}

The Katzman Automatic Imaging Telescope (KAIT), also at the Lick Observatory
 \citep{2003PASP..115..844L}, responded to the trigger and began taking
 unfiltered exposures starting 42 seconds after the trigger time. This paper
 includes 206 unfiltered
 data points, which have been reduced in a standard way and then calibrated
 to the $R$-band \citep{2003PASP..115..844L}. These data, once calibrated, are
 shown along with the PAIRITEL data in Figure \ref{KP}.\par
The first KAIT data point has a central time of 57 seconds, with a total
 exposure time of 30 seconds. Given the highly variable nature of early time
 GRB emission, the large error bars on the value, the long duration over which
 the magnitude was measured and (as later discussed) its outlier nature,
 we felt that this magnitude did not provide a useful measure the $R$-band
 emission over this time. We therefore excluded it from the later analysis.\par

\subsection{Gemini}

Our last optical data were acquired with Gemini-North using the
Gemini Multi-Object Spectrograph (GMOS) in imaging mode with the $r$-band
filter. The observations began at 10:22UT on 19th March 2008 and consisted
of $5 \times 150$ second exposures.  The data were reduced using the
{\sc gemini-gmos} routines within {\sc iraf}.  No significant flux was
detected at the location of the afterglow, as reported in Table A5.

\subsection{Polarization}

Three linear polarimetric observation sets were carried out with the ESO-VLT
 during the late afterglow evolution as shown in Table\,\ref{tab:pol}. The
 observations allowed us to derive rather stringent upper limits although
 still compatible with past late-time afterglow detections
 (\citealt{2005AIPC..797..144C}, see also, however,
 \citealt{2003ApJ...583L..63B}) and substantially lower than the early time
 afterglow measurement by \citet{2009Natur.462..767S} for GRB~090102.
 In the case of GRB~090102, however, the detection is taken at 160 seconds;
 a time when \citet{2009Natur.462..767S} argue that the flux should be
 dominated by reverse shock emission.\par
The detection of linear polarization at the level of a few percent in the
 light from GRB optical afterglow is well within the prediction of the
 external shock scenario (\citealt{2004IJMPA..19.2385Z}, and references
 therein) and indeed it is still one of the most relevant observational
 findings supporting it (e.g., \citealt{2010xpnw.book..215C}). On the other
 hand, a comprehensive framework predicting the polarization degree and
 position angle evolution during the afterglow had to deal with the increasing
 complexity and variety of behaviours shown by the afterglow population. In
 general, the late-afterglow optical polarization is related to three main
 ingredients: the emission process able to generate highly polarized photons
 (i.e., synchrotron), the ultra-relativistic motion and the physical beaming of
 the outflow \citep{1999MNRAS.309L...7G,1999ApJ...524L..43S}. Therefore the
 polarization time-evolution is in principle a powerful diagnostic of the
 afterglow physics, and many attempts were carried out to compare observations
 to models (e.g., \citealt{2003A&A...410..823L,2004A&A...422..121L}) with
 particular emphasis to the jet structure \citep{2004MNRAS.354...86R}.\par
During the polarimetric observations of  GRB~080310 the afterglow showed a
 smooth decay (see Figure \ref{all_lc}) without any detectable temporal
 break or spectral change. Polarization below $\sim 2.5$\% cannot put specific
 constraints on the afterglow modeling or the jet structure. These results,
 are however compatible with what would be expected at late times, as this is
 when the forward shock should dominate emission and the magnetization signal
 of the fireball is lost in the interaction with the surrounding medium.\par
\begin{table}
  \centering
  \caption{ESO-VLT polarimetric observations of the late afterglow of
 GRB~080310}
  \label{tab:pol}
  \begin{tabular}{cccc} 
    \hline
    $t-t_{0}$ (s) & $T_{exp}$ (s) & Polarization (\%) & Band \\
    \hline
    87171 & 1447 & $< 2.5$ & $V$ \\
    169501 & 1447 & $< 2.5$ & $V$ \\
    253724 & 4607 & $<2.6$ & $V$ \\
    \hline
  \end{tabular}
\end{table}

\subsection{Observations from literature}

Data from the Palomar 60 inch telescope \citep{2006PASP..118.1396C} were
 obtained from the Palomar 60 inch-\swift\ Early Optical Afterglow Catalog
 \citep{2009ApJ...693.1484C}, in which the 29 GRBs between the
 1$^{\textrm{st}}$ of April 2005 and the 31$^{\textrm{st}}$ of March 2008 with
 P60 observations beginning within the first hour after the initial
 \textit{Swift}-BAT trigger are presented.\citet{2009ApJ...693.1484C} reduce
 data in the {\sc iraf} environment, using a custom pipeline detailed in
 \citet{2006PASP..118.1396C}.  Magnitudes were calculated using aperture
 photometry and calibration performed using the USNO-B1
 catalog\footnote{http://www.nofs.navy.mil/data/fchpix} and the data were
 corrected for dust extinction using the extinction maps of
 \citep{1998ApJ...500..525S}.\par
Further published data for GRB~080310 were obtained from
 \citet{2010ApJ...720.1513K}, in which SMARTS-ANDICAM data are detailed as part
 of an extensive survey of optical data for GRBs in both the pre-\swift\ and
 \swift\ eras. \citet{2010ApJ...720.1513K} reduce their data using standard
 procedures in {\sc iraf} and {\sc midas}. Both aperture and PSF photometry
 were used in the derivation of magnitudes, when comparing to standard
 calibrator stars.\par
The 0.6m Super-LOTIS (Livermore Optical Transient Imaging System) telescope,
 located at the Steward Observatory (Kitt Peak, Arizona;
 \citealt{2004AN....325..667P}) began $R$-band observations of the error
 region of GRB~080310 at 08:38:43 UT, 44 seconds
 after the start of the burst \citep{2008GCN..7387....1M}. The OT detected by
 \citet{2008GCN..7381....1C} and confirmed by \citet{2008GCN..7382....1C} was
 not apparent in the initial images, even when stacking the first three 10
 second exposures. However, the subsequent 20 second exposures do show the
 optical transient without stacking, which suggests that the GRB brightened
 in the $R$-band during the first two minutes after detection. A nearby USNO-B
 star was used to derive the $R$ magnitude.\par

\section{Modelling}

To model the emission of GRB~080310 we begin by fitting the BAT and XRT data,
 before then extending the model into the lower energy bands.\par

\subsection{Initial modelling of the high energy emission}

We expand on the previous work done by \citet{2010MNRAS.403.1296W}, where a
 sample of 12 GRBs were selected and fitted using the pulse model of
 \citet{2009MNRAS.399.1328G}. In this model, the prompt emission component of
 GRB emission is split into a series of pulses, where each pulse is considered
 to be the result of a relativistically expanding thin spherical shell that
 emits isotropically. It was assumed that each of the pulses was in the fast
 cooling regime \citep{1998ApJ...497L..17S} and that each of the X-ray and
 $\gamma$-ray spectra could be fitted with a temporally evolving Band function
 \citep{1993ApJ...413..281B}. A Band spectral energy distribution is a
 smoothly broken power-law. Below, we show a modified Band function, which
 includes temporal evolution.\par
\begin{equation}
B(z)=B\left\{
\begin{array}{ll}
z^{b_{1}-1}e^{-z}, & \mbox{$z$ $\le$ $b_{1}-b_{2}$,} \\
z^{b_{2}-1}\left(b_{1}-b_{2}\right)^{b_{1}-b_{2}}e^{-\left(b_{1}-b_{2}\right)}, & \mbox{$z$ $>$ $b_{1}-b_{2}$,} \\
\end{array}
\right.
\label{bandfnc1}
\end{equation}
where
\begin{equation}
z=\left(\frac{E}{E_{f}}\right)\left(\frac{T-T_{ej}}{T_{f}}\right)^{d}.
\label{zdef}
\end{equation}
The times in Eqn \ref{zdef} are all in the observed frame, where $T_{ej}$ is
 the time of shell ejection from the central engine and $T_{f}$ is the time
 at which the last photon arrives from the shell along the line of sight. $B$
 is the normalisation to the Band function, with $b_{1}$ and $b_{2}$ being the
 low and high energy photon indices of the Band function, respectively. The
 spectral shape of each pulse evolves in time as determined by the temporal
 index $d$. In \citet{2010MNRAS.403.1296W} $d=$ -1 due to the assumption that
 the emission process is synchrotron under the standard internal shock model.
 It is this value of temporal index that was also used by
 \citet{2009MNRAS.399.1328G} in the original derivation of the pulse profiles,
 meaning that only a value of $d=$ -1 is strictly consistent with the
 original pulse model.\par
The prompt pulse model describes the morphology of the prompt light curve and
 the rapid decay phase as observed in the high-energy bands. In addition
 to this, an afterglow component was also included in the modelling as
 outlined in \citet{2007ApJ...662.1093W}. The afterglow component has a
 functional form as outlined in Eqn \ref{orig_ag}, which comprises an
 exponential phase that transitions into a power-law decay.\par
\begin{equation}
f_{a}\left(t\right)=\left\{
\begin{array}{ll}
F_{a}\exp\left(\alpha_{a}-\frac{t\alpha_{a}}{T_{a}}\right)\exp\left(\frac{-T_{r}}{t}\right), & \mbox{$t$ $<$ $T_{a}$,} \\
F_{a}\left(\frac{t}{T_{a}}\right)^{-\alpha_{a}}\exp\left(\frac{-T_{r}}{t}\right), & \mbox{$t$ $\geq$ $T_{a}$.} \\
\end{array}
\right.
\label{orig_ag}
\end{equation}
In Eqn \ref{orig_ag}, $f_{a}\left(t\right)$ is the flux from the afterglow
 at time $t$, $F_{a}$ gives the flux at the transition time between the
 exponential and the power-law components, $T_{a}$. $T_{r}$ is the rise time
 and finally $\alpha_{a}$ is the index that governs the temporal decay of
 the power-law phase.\par
Combining the prompt pulses and the afterglow component we adopted the
 same method of fitting the data from the \swift\ BAT and XRT instruments
 as \citet{2010MNRAS.403.1296W}, by first identifying the individual pulses 
 in the BAT light curve and allowing their parameters to be fitted by
 minimizing the $\chi^{2}$ fit statistic for both the BAT and XRT light curves.
 The afterglow component was then fitted by allowing the routine to find the
 optimum values for the characteristic times and normalizing flux shown in
 Eqn \ref{orig_ag}.\par
\begin{figure*}
\begin{center}
\includegraphics[height=17.5cm]{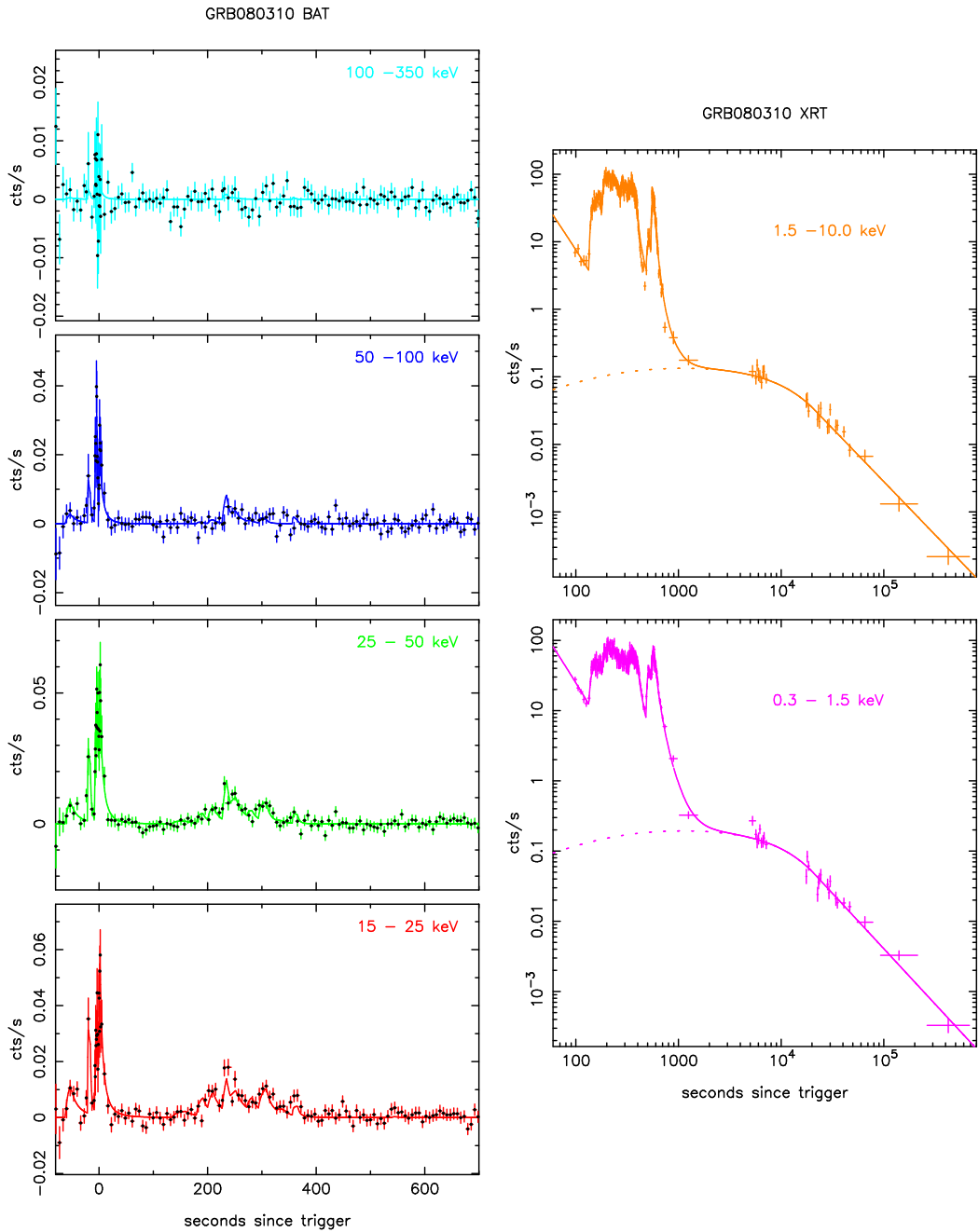}
\end{center}
\caption{BAT and XRT data, showing the fitted \citet{2010MNRAS.403.1296W}
 model. Left panels: BAT bands (Top to bottom: 100-350 $\keV$,
 50-100 $\keV$, 25-50 $\keV$ \& 15-25 $\keV$). Right panels: XRT bands (Top:
 XRT hard band, bottom: XRT soft band). Pulse parameters have
 been fixed, whilst the afterglow parameters have been fitted. Solid lines show
 the total fit (pulses and afterglow), whilst dashed lines show only the
 afterglow component to the fit.}
\label{batxrt}
\end{figure*}

The simultaneous BAT and XRT fit is plotted in Figure \ref{batxrt} and shows
 several important characteristics of both the burst and the model. Firstly,
 there is a lot of structure evident in the light curves, particularly during
 the first 1000 seconds of the XRT light curves. In this fit, sixteen unique
 pulses have been identified and can be seen to fit the XRT data accurately.
 In the original fit from \cite{2010MNRAS.403.1296W} there were only ten
 pulses, however this failed to fully model some of the structure in the
 softest BAT band (15-25 $\keV$) between 200 and 400 seconds. The additional
 pulses now provide a better fit to this time, across all the BAT and XRT
 bands with a reduced $\chi^{2}$ statistic of 1.68 for 1212 degrees of
 freedom. The quality of this fit is formally not statistically acceptable,
 however the largest contribution to the $\chi^{2}$ value is from small
 scale intrinsic fluctuations in the data. The properties of the sixteen pulses
 are listed in Table \ref{pulsetable1}.\par

\begin{table}
 \centering
  \caption{Properties of the prompt pulses identified in the BAT and XRT
 light curves. These include the peak time ($T_{pk}$), peak energy at this time
 ($E_{pk}$), the rise time ($T_{r}$) and arrival time of the last photon
 ($T_{f}$) for each pulse.}
  \label{pulsetable1}
  \begin{tabular}{@{}cccccc}
    \hline
    Pulse & $T_{pk}$ (s) & $E_{pk}$ (keV) & $T_{r}$ (s) & $T_{f}$ (s) &
 $b_{1}$  \\
    \hline
    1 & -52.8  & 200 & 9.7 & 44.5 & -1.49 \\
    2 & -16.0 & 200 & 5.0 & 6.0 & -1.20 \\
    3 & -4.6 & 200 & 3.8 & 11.7 & -0.40 \\
    4 & 1.8 & 200 & 2.7 & 17.2 & -1.50 \\
    5 & 159.0 & 12.3 & 25.3 & 74.0 & -0.30 \\
    6 & 191.6 & 13.4 & 12.4 & 39.8 & -0.02 \\
    7 & 210.0 & 21.3 & 10.0 & 46.7 & -0.16 \\
    8 & 235.0 & 58.0 & 8.0 & 24.4 & -0.13 \\
    9 & 251.8 & 42.0 & 10.0 & 52.4 & -0.20 \\
    10 & 282.0 & 15.8 & 10.0 & 40.0 & -0.10 \\
    11 & 308.7 & 16.1 & 15.0 & 32.5 & 0.24 \\
    12 & 342.7 & 7.8 & 15.0 & 40.1 & -0.18 \\
    13 & 366.0 & 34.3 & 10.0 & 11.5 & -0.38 \\
    14 & 390.0 & 1.2 & 23.0 & 56.1 & 0.21 \\
    15 & 513.8 & 3.8 & 31.1 & 193.8 & -1.26 \\
    16 & 582.1 & 2.4 & 43.9 & 66.5 & -0.09 \\
    \hline
\end{tabular}
\end{table}

\subsection{Extrapolating to the optical and IR}

Previously, no optical or IR data have been included when modelling the
 light curve and a fit to only the \swift\ BAT and XRT data has been produced
 (Figure \ref{batxrt}). However, given the rapid response of optical and IR
 instruments to the trigger for GRB~080310 we extend the model to include
 these new sources of data.\par
The simplest approach to fitting the optical and IR light curves for GRB~080310
  was to use the fitting routine from \citet{2010MNRAS.403.1296W} and simply
 extrapolate the Band functions for both the pulses and also the afterglow to
 these lower energies. In this initial attempt all of the parameters
 previously discussed were held at the values obtained for the high energy
 fit, to see what modifications might be necessary to both components.\par
\begin{figure*}
\begin{center}
\includegraphics[height=15cm,angle=270]{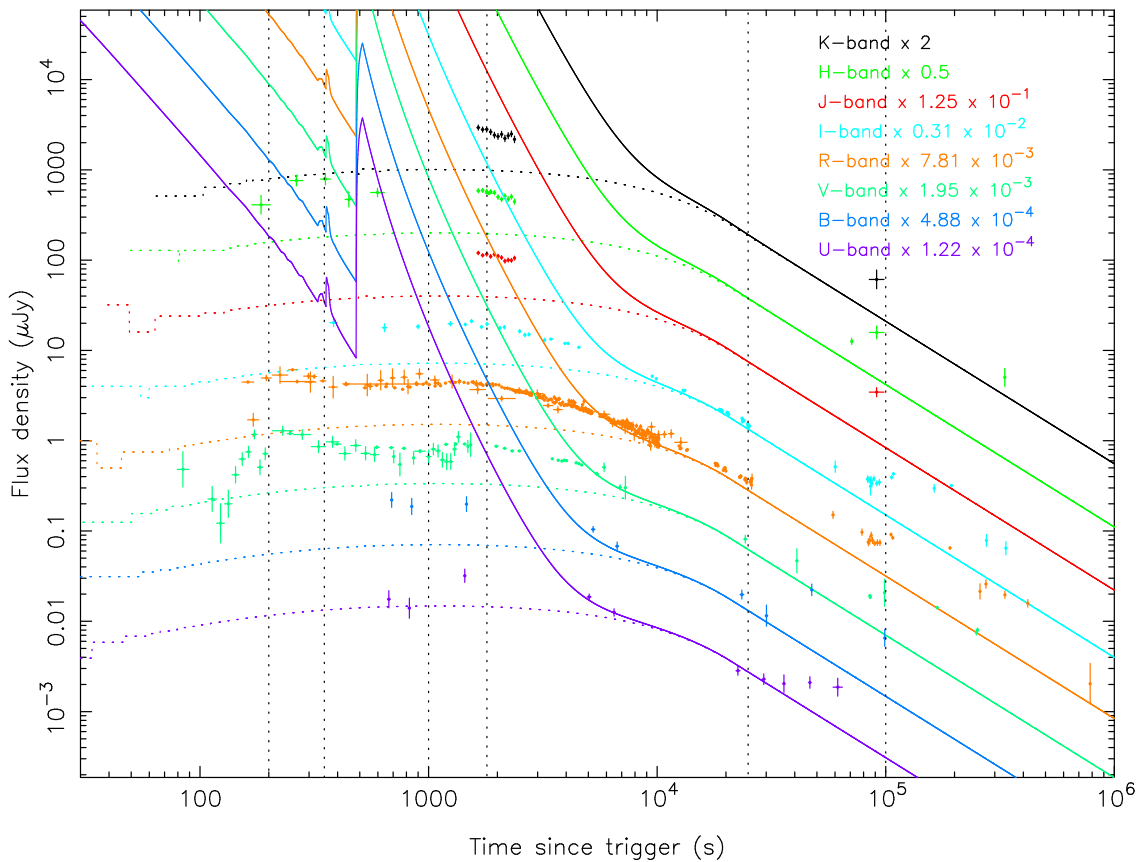}
\end{center}
\caption{Optical and IR light curves obtained by extending the high energy fit
 to the prompt pulses and afterglow component.The spectral shape of the pulses
 is assumed to be a Band function extrapolated back to the optical and IR data
 ($K$, $H$, $J$, $I$, $R$, $V$, $B$ and $U$ bands). Vertical dashed lines
 denote times of interest later discussed, at which the SEDs are considered at
 in more detail. The coloured dashed lines show the afterglow component to the
 extrapolation for each band.}
\label{bandfit}
\end{figure*}
Whilst it provides an acceptable fit to the XRT and BAT bands, the original
 pulse model vastly over predicts the optical and IR fluxes from the pulses
 (Figure \ref{bandfit}),
 which implies there must be a break in the pulse spectrum between the X-ray
 and the optical and IR energies. Such a break is expected in a synchrotron
 spectrum, but can also been seen in thermal spectra as the Rayleigh-Jeans
 tail. Additionally, the afterglow prediction from the BAT and XRT fit is
 not consistent with the optical data at late times where the prompt
 component to the light curve is negligible. Figure \ref{bandfit} also shows
 that the temporal decay index of the power-law phase of the afterglow gives
 rise to a decay which is more rapid than the optical data suggest.\par
An alternative method of reducing optical flux is to invoke dust absorption,
 however, as reported in \citet{2008GCN..7406....1P}, extinction due to dust
 is low at $A_{V}~=~0.10~\pm0.05$ at an average time of $T_{0}(+1,750)$ seconds
 for GRB~080310. As not only the late time emission seems unaffected by such
 absorption, but also by 2,000 seconds after trigger, we chose to favour a low
 energy spectral break in our modelling of the optical emission. The
 following sections of this paper detail the implementation of such
 alterations to the \citet{2010MNRAS.403.1296W} model.\par

\subsection{Modifications to the model}

\subsubsection{An additional spectral break}

\begin{figure}
\begin{center}
\includegraphics[height=8cm,angle=270]{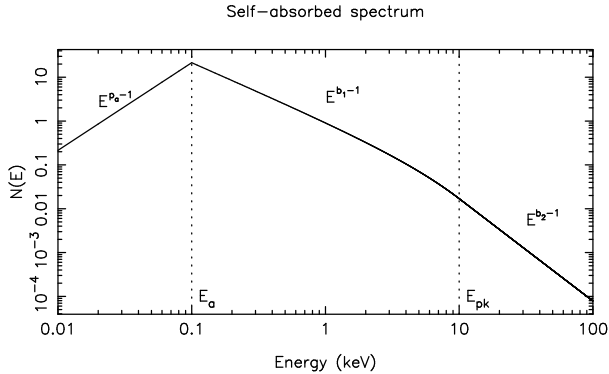}
\end{center}
\caption{Example modified spectrum including additional spectral break.}
\label{sa_1}
\end{figure}

To reduce the flux from each pulse in the optical and IR light curves of
 GRB~080310 we introduced an additional break to the spectrum for each
 prompt pulse. An example of such a spectrum is shown in Figure \ref{sa_1},
 where a regular Band function has a low energy break below $E_{pk}$. To
 fully describe this break, three parameters are needed; a value of the
 break energy ($E_{a}$), the spectral index of the power-law slope in the
 low energy regime ($p_{a}$) and a temporal index which describes how the
 break energy evolves in time ($d_{a}$). The value of $E_{a}$ is defined at
 $T_{pk}$, when the emission from the pulse is at a maximum. The entire
 pulse spectrum already evolves in time, according to the index $d$ as
 shown in Eqn \ref{zdef}, and so the expectation was that the time evolution
 of $E_{a}$ would be related. Previously a value of $d=$ -1 was used, which
 when integrating of the Equal Arrival Time Surface (EATS) means a pure
 Band function can be recovered in the source frame given a Band function
 in the observed frame. Other values of $d$ prevent assumptions to be made
 concerning the source spectrum. We allowed the index $d_{a}$ to
 be fitted independently, to test whether the evolution of the break was
 the same as the rest of the spectrum. Again, we note that only $d_{a}=$ -1
 is fully consistent with the original \citet{2009MNRAS.399.1328G} pulse
 model. The functional form of the new spectral model for each pulse is
 shown in Eqn \ref{safunc}.\par
\begin{equation}
N(E)=\left\{
\begin{array}{ll}
A E^{p_{a}-1} \exp \left(\frac{-E_{a}}{E_{c}}\right)E_{a}^{-\alpha}, & \mbox{ $E$ $\leq$ $E_{a}$,}  \\
A E^{b_{1}-1} \exp \left( \frac{-E}{E_{c}} \right), & \mbox{ $E_{a}$ $<$ $E$ $\leq$ $E_{pk}$,} \\
A E^{b_{2}-1} \exp \left( \alpha - \beta \right) E_{pk}^{-(\alpha - \beta)}, & \mbox{ $E$ $>$ $E_{pk}$.} \\
\end{array}
\right.
\label{safunc}
\end{equation}
The model spectrum for photon emission assumes several things. Firstly,
 there is a single underlying population of relativistic electrons, whose
 energies can be described by a broken power law such as that described in
 \citet{2009MNRAS.398.1936S}. Such a spectrum of electron energies leads
 to a similar photon spectrum: a singly broken power-law, which corresponds
 to the observed Band function. However, it would be unphysical for this
 spectral shape to extend indefinitely to low energies, particularly as
 the electron energy spectrum has an associated minimum energy $E_{m}$. This
 minimum electron energy has a related emission energy at a frequency
 $\nu_{m}$. With no self-absorption, the spectral index to the emission
 spectrum changes to $\frac{1}{3}$ at photon frequencies below $\nu_{m}$,
 but if self-absorption is present then at low energies we would expect to
 see a steeper spectral index of 2 if the absorption frequency is less than
 $\nu_{m}$ or alternatively a spectral index of $\frac{5}{2}$ in the case of
 $\nu_{m}~<~\nu_{a}$. In the former case, if there is an intervening spectral
 range between $\nu_{a}$ and $\nu_{m}$, a spectral index of $\frac{1}{3}$ is
 expected between these two frequencies.\par
To keep the model simple, due to the limited nature of the spectral
 information available, we only included a single additional break in the
 spectrum. By allowing the resultant spectral index to be constrained by the
 observations, rather than assuming a value, we hope to better understand the
 physics required to explain the spectrum of prompt GRB pulses.\par

\subsubsection{The afterglow}

As seen in Figure \ref{bandfit}, prompt pulses without a low energy break
 significantly over predict
 the observed flux in the optical and IR light curves prior to 10,000 seconds.
 By introducing a low energy break to the pulse spectra, we hoped to
 completely remove the prompt contribution to the total emission, and model
 the entire light curve with afterglow emission. One reason for considering
 this is the smooth nature of the optical light curves. Whilst there is
 some variation during the plateau seen between 200 and 3,000
 seconds, this is at a low level, and happens smoothly over a large period
 of time. If the prompt pulses observed at higher energies were also the
 dominant source of emission at these energies, then we might expect to see
 similar structure in the optical and IR light curves to that shown in Figure
 \ref{batxrt}.\par
The first issue to address with the afterglow fit was to account for the
 small difference in temporal decay index ($\alpha_{a}$) between the higher
 and lower energy bands. To do so, we included an additional parameter that
 described the difference in this index, and allowed it to be fitted. By
 including this, we could account for the slightly shallower decay of the
 power-law phase in the optical and IR channels. The change in $\alpha_{a}$
 required was found to be small, at a value of approximately 0.2, but the
 associated errors in each instance showed it to be inconsistent with zero.
 This may be explained physically by a slight curvature in the spectrum of
 the afterglow between the optical and X-ray bands. This, however, is not
 a large enough difference to suggest a spectral break, such as the one
 introduced to the prompt pulses. The GRB afterglow flux is also thought to
 be synchrotron emission. However, at these late times the radiation is
 usually assumed to come from an optically thin plasma, and so a
 self-absorption break would be expected at energies lower than the optical
 bands observed for GRB~080310. We also ruled out contamination in the
 optical and IR wavebands as being the source of this difference as any
 emission from the host should be constant with time. This should add a
 constant offset to the data, and should the afterglow emission reach a
 comparable order of magnitude, a plateau at the end of the optical decay
 would be expected. This plateau is not observed, and the difference between
 the low and high energy temporal decay indices of the afterglow is also
 determined by data prior to when the host would be seen to make a
 significant impact on the optical and IR light curves.\par
With no prompt component capable of rising quickly, we also had to modify
 the manner in which the afterglow rises, to account for the rapid increase
 in flux seen in the V-band data at approximately 100 seconds in Figure
 \ref{all_lc}. To do this, we introduced a third part to the functional form
 that is shown Eqn \ref{orig_ag}. This extra regime is shown in the top line
 of Eqn \ref{ageqn}, and describes a power-law rise in flux, with an index
 of 2. We also allowed the afterglow to be launched at a time that was
 independent to the trigger time of the GRB, by introducing a launch time
 ($T_{l}$), which offsets the afterglow in time to better fit the timing of
 the rise observed in the V-band. These modifications are shown in Eqn
 \ref{ageqn} and are used for the afterglow-dominated modelling only. In the
 case of the prompt dominated fit, we return to the afterglow model of
 \citet{2007ApJ...662.1093W} as explained in \S 3.5.\par
\begin{equation}
f_{a}(t)=\left\{
\begin{array}{ll}
F_{a}\exp\left(\alpha_{a}-\frac{(T_{r}-T_{l})\alpha_{a}}{T_{a}-T_{l}}\right)\left(\frac{t-T_{l}}{T_{r}-T_{l}}\right)^{2}, & \mbox{ $t$ $\leq$ $T_{r}$,} \\
F_{a}\exp\left(\alpha_{a}-\frac{(t-T_{l})\alpha_{a}}{T_{a}-T_{l}}\right), & \mbox{ $T_{r}$ $<$ $t$ $\leq$ $T_{a}$,} \\
F_{a}\left(\frac{t-T_{l}}{T_{a}-T_{l}}\right)^{-\alpha_{a}}, & \mbox{ $t$ $>$ $T_{a}$.} \\
\end{array}
\right.
\label{ageqn}
\end{equation}

\subsection{Afterglow dominated fit}

Given the smooth nature of the optical and IR light curves, our initial use
 of the low-energy spectral break in the prompt pulse spectrum was to remove
 the prompt component to the optical light curves entirely, and try and
 explain the early time behaviour with only afterglow emission. To do this, the
 prompt pulses were initially switched off from the optical and near infrared
 fit. With these removed, the afterglow parameters $T_{l}$, $T_{r}$, $T_{a}$,
 $F_{a}$, $\alpha_{a}$ and the change between optical and high energy values
 of $\alpha_{a}$ were fitted. Having obtained values for these parameters
 using the fitting routine, we re-introduced the prompt pulses, allowing
 $p_{a}$ to be fitted, to find the minimum
 break in the pulse spectra required to remove the prompt optical and IR
 flux. Where the fit statistic tended to unphysical values of $p_{a}$ we
 checked the effects of forcing $p_{a}$ to be $\frac{5}{2}$. When this was
 undertaken, we found that whilst there was an increase in the reduced
 $\chi^{2}$ statistic, this was incredibly small,  approximately
 $1 \times 10^{-3}$,for 1642 degrees of freedom.\par

\subsubsection{Light Curves}

\begin{figure*}
\begin{center}
\includegraphics[height=15cm,angle=0]{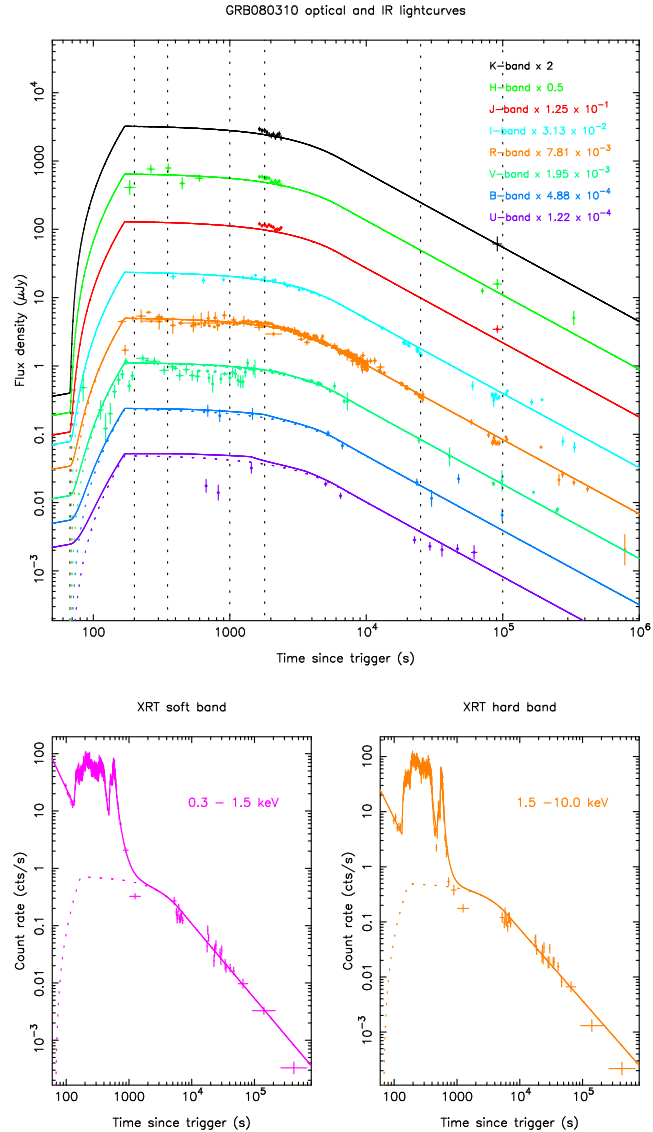}
\end{center}
\caption{Simultaneous fit to optical, IR and XRT data, in which the
 early time optical and IR are fitted by the afterglow emission. Top panel:
 Optical and IR bands fitted with low energy break ($K$, $H$, $J$, $I$, $R$,
 $V$, $B$ and $U$ bands). Bottom left: XRT soft band fit. Bottom right: XRT
 hard band fit. The afterglow parameters are fitted, as are values for
 $d_{a}$, whilst $E_{a}=$ 0.3 keV and $p_{a}$ has been set to $\frac{5}{2}$.
 Coloured dashed lines show the afterglow component in each band. The
 reduced $\chi^{2}$ for this fit is 2.44 with 1,642 degrees of freedom.}
\label{fit_ag}
\end{figure*}
Figure \ref{fit_ag} shows the light curves for the best fit found using the
 method described above, for which a reduced $\chi^{2}$ value of 2.44 with
 1,642 degrees of freedom was found. The late time XRT afterglow isn't as well
 fitted as in the original \citet{2010MNRAS.403.1296W} method (Figure
 \ref{batxrt}), where only the high energy bands were included. This is
 particularly the case for the last data points, and those at the transition
 between the prompt and the afterglow phase in the X-ray bands. The
 cause of this is due to the greater number of optical points being the
 largest constraint on the afterglow parameters, particularly the
 characteristic times, $T_{a}$ and $T_{r}$. Because these times have changed
 from the original fit, the temporal decay index in the power-law phases has
 reduced from $1.58^{+0.04}_{-0.03}$ to $1.30^{+0.01}_{-0.01}$. The afterglow
 parameters all have a reduced error in this newer fit as the optical and IR
 data points allow for more accurate fitting with a greater quantity of data.
 The characteristic times also have improved accuracy, with $T_{r}$ =
 $223^{+16}_{-20}$ seconds, $T_{a}$ = $6,368^{+495}_{-50}$ seconds and $T_{l}$
 = $120^{+9}_{-11}$ seconds. The errors quoted in this instance were obtained
 when floating all of the parameters mentioned simultaneously. Once the
 afterglow parameters were in place, the final bulk value of $p_{a}$ was
 found to tend to $5.6$, with a 1$\sigma$ lower limit of 5.5 and an
 unconstrained upper limit. The lower limit is tightly constraining, and
 likely due to the fitting routine trying to remove contributions from the
 brightest early pulses, which therefore affects the value assigned to the
 weaker pulses as well. The lack of an upper limit suggests that the fitting
 routine is attempting to remove the prompt component entirely from the
 optical emission, as driven by the morphology of the light curve particularly
 the dense sample of KAIT data points. The value of 5.6 is very
 significantly steeper than that expected at either 2 or $\frac{5}{2}$. The
 suggested lower limit from the $\chi^{2}$ distribution is surprisingly tight,
 given the shape of the curve obtained for the statistic when only varying
 $p_{a}$ as the distribution is remarkably flat for a broad range of values.
 Looking at the fit statistic by eye, it would appear that the expected values
 from synchrotron or thermal spectra are not entirely ruled out, but lie within
 the lower limits of where the fit is a faithful representation of the
 data. To test this, we forced $p_{a}$ to a value of $\frac{5}{2}$ for all the
 prompt pulses, and noticed that the value for the $\chi^{2}$ statistic
 increased by a nearly negligible amount. By using $p_{a}$ = $\frac{5}{2}$
 self-absorption becomes a viable explanation for the physical mechanism
 causing the observed level of optical and near infrared flux. It is therefore
 this value that was adopted for $p_{a}$ for the prompt pulses when trying to
 suppress them at early times in the lower energy regimes.\par
The broad scale morphology of the optical and NIR light curves are described by
 the fast rising afterglow, with the rise observed in the $V$-band being
 fitted and the level of flux of the plateau seen between 100 and 1,000
 seconds being consistently modelled in five of the six bands in which it can
 be seen. The exception to this is that the $U$-band model light curve
 morphology seems to over-predict the emission actually observed in the first
 1,000 seconds, which may be a result of the correction required to removed the
 effects of absorption from the Lyman forest as described in the section
 detailing the UVOT observations. The $I$-, $R$- and $V$-band data
 during this plateau phase do appear to have a slight dip in flux, which is
 not picked up by the afterglow fit. Additionally, the first $V$-band points
 indicate a decline in flux prior to the launching of the afterglow. To fit
 this, one of the early time pulses would have to be switched on at low
 energies, implying that at the earliest times the prompt component is still
 important in the optical and IR regimes.\par
The $V$-band data between 100 and 1,000 seconds are from three sources, VLT
 $V$-band, UVOT $v$-band and the UVOT \textit{white} filter, which is
 normalised to the $v$-band. This normalisation is done at a time when there
 is near simultaneous UVOT $v$-band and \textit{white} filter coverage. The
 \textit{white} data are then adjusted at this time, to make the observed flux
 the same as that in the $v$-band. The same correction factor is then applied
 to all the \textit{white} filter points, which relies on an intrinsic
 assumption that the spectrum is not varying between the $v$-band and
 \textit{white} filter for all observations. If the total GRB low
 energy emission is primarily from the afterglow, this assumption is accurate,
 as the spectral break in the Band function will not migrate to the optical
 part of the spectrum. We noted, upon inspection of the data, that the UVOT
 points between 300 and 2,000 seconds appear to not be entirely consistent with
 the near simultaneous VLT data. The VLT flux values have much lower
 associated error, which led to us reprocess the UVOT data, to verify the
 values. Doing this we obtained the same UVOT fluxes. To account for the
 discrepancy, we introduced a systematic error in the UVOT points at the level
 of 10\% of the measured flux for each datapoint. Analysis of only the VLT
 data, still shows the deficit of flux, which we discuss throughout this work,
 at a level that is significant given the small errors associated with the
 data. As this feature is significant in instruments other than UVOT, we
 believe that the discrepancy from the UVOT instrument does not impinge on the
 main conclusions of this work.\par
Figure \ref{all_lc} would suggest
 that the decline in flux is not marked by a change of instrument.\par
The temporal index $d_{a}$ governing the time evolution of $E_{a}$ is tightly
 constrained in this model, to a value of $d_{a}$ = $-1.01^{+0.05}_{-0.03}$.
 Not only does this match the temporal evolution of the characteristic energy
 $E_{c}$ in \citet{2009MNRAS.399.1328G}, but the 1-$\sigma$ errors quoted are
 restricting. Having analysed the one dimensional $\chi^{2}$ surface, it is
 clear that a more rapid evolution of the temporal index allows the prompt
 pulses to peak at earlier times, when the normalising flux at the peak of
 each pulse is still significant, causing an over-prediction of early time
 flux. At lower values the $\chi^{2}$ distribution is fairly flat, as the
 prompt flux normalisation is so low, it cannot be seen to contribute to the
 total light curve.\par

\subsubsection{Spectral Energy Distributions}

\begin{figure*}
\begin{center}
\includegraphics[height=20.0cm,angle=0]{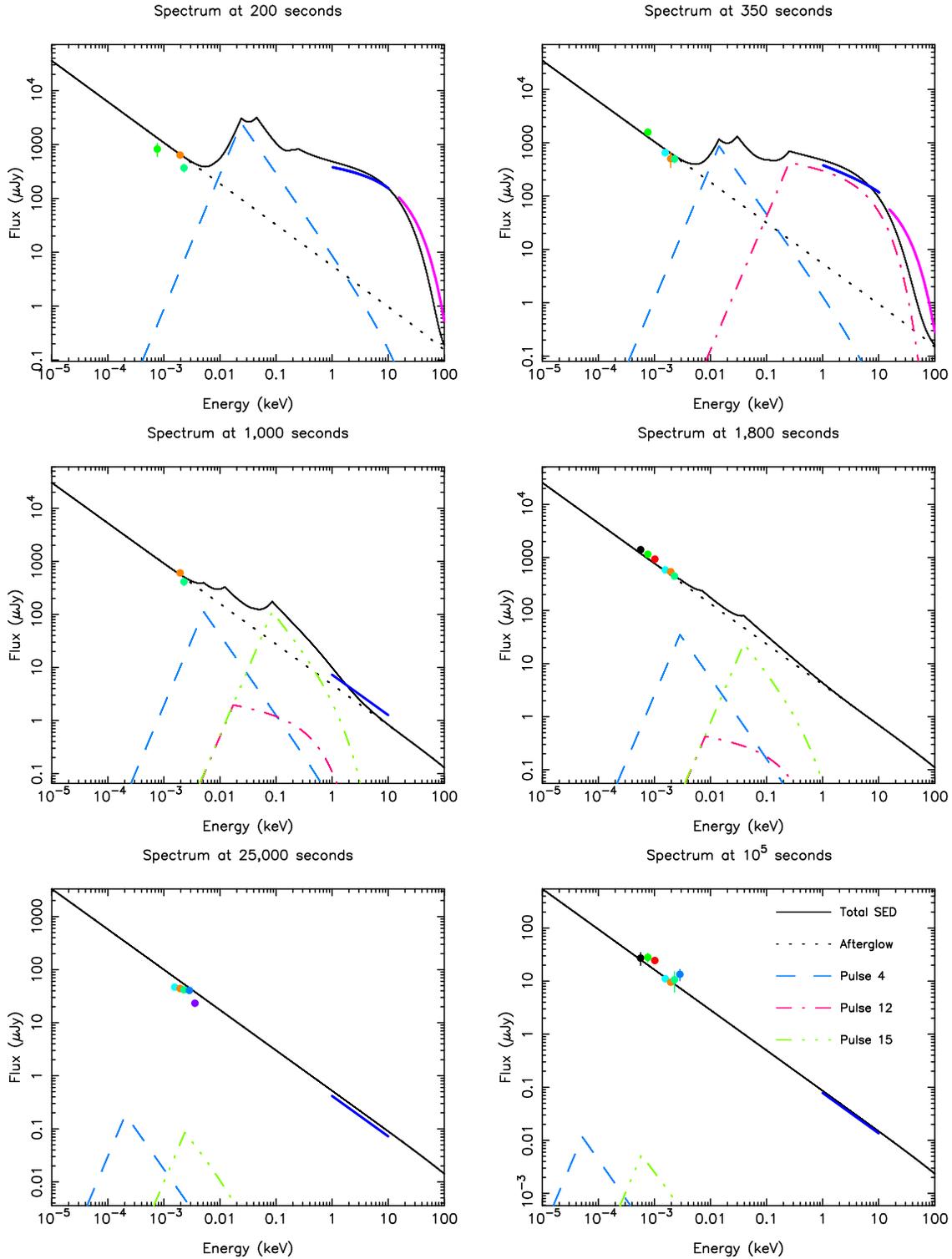}
\end{center}
\caption{SEDs for fiducial times in the light curve of GRB~080310 using the
 afterglow model to fit the early time, low energy emission. Prompt pulses are
 included with a bulk value of $p_{a}$ = 2.5. The solid line is the total SED
 and the dotted line is the contribution from the afterglow component. Pulses
 4, 12 and 15 are also shown by the dashed and dotted lines as outlined in the
 key shown in the bottom right panel. These are included as examples to show
 the evolution of some of the longer lasting pulses in the fit. The colours of
 the data points follow the same scheme as that used in the light curves, with
 the BAT and XRT data being represented by lines which extend along their
 entire coverage.}
\label{sed1}
\end{figure*}

Another way of comparing the model to the observed data is to consider the
 spectral energy distribution (SED) at fiducial times. The times selected are
 indicated in Figures \ref{bandfit}, \ref{fit_ag} and \ref{fit_prompt} as
 dashed vertical lines. These correspond to times of 350 seconds (where there
 are early-time optical data and the activity in both the X-ray and
 $\gamma$-ray bands), 1,800 seconds (a time where the high energy only model
 shown in Figure
 \ref{bandfit} suggests the afterglow is becoming significant), $2.5$ $\times$
 $10^{4}$ and $1$ $\times$ $10^{5}$ seconds (both of which are later times
 that should be dominated by the afterglow). Figure \ref{sed1} shows the model
 SED at these times, and also those at both 200 and 1,000 seconds. The times
 shown were chosen to both allow for comparison with the maximum number of
 bands (both optical and X-ray) at each time and to show times where the SED
 can discriminate between prompt and afterglow emission at optical and IR
 energies.The SEDs shown in Figure \ref{sed1} are those for the fit where the
 afterglow dominates the entire low energy emission. In each case, error bars
 are plotted, but in some panels are too small to be seen.\par
Figure \ref{sed1} shows that by combining the prompt pulses and the afterglow
 component, the model SEDs have complicated structure at high energies. The
 first panel shows the SED at 200 seconds, a time by which the first six pulses
 have peaked. As an example of time evolution, the spectral contribution
 from pulse four is shown with the dashed line in this and subsequent panels.
 The temporal evolution of pulses 4, 12 and 15 can be traced through
 all six panels. At 200 seconds pulse 4 is the one of the three pulses causing
 deviation from the afterglow Band function (the others being pulses 1 and 6).
 Following it through all six panels of Figure \ref{sed1} shows the general
 evolution of all the pulses. The peak flux reduces with time, and the energy
 at which this peak is seen becomes lower with increasing time also. By the
 late panels, the component of pulse 4 can still be seen on the axes, however,
 being several orders of magnitude lower than the afterglow Band function, it
 does not produce an observable feature in the total SED. Because the spectral
 index for all the pulses is at a value of $p_{a}$ = 2.5, before the peaks
 migrate to the optical regime, their contribution to the total flux at these
 energies is negligible. Once the peak has had time to evolve to such energies,
 the normalizing flux has been diminished, so the pulse contributions remain
 negligible.\par
Figure \ref{sed1} also shows that the dominant pulses in the SED of GRB~080310
 do not have to be those most recently launched. Pulse 12 is emitted after
 pulse 4, as shown in the first two panels of the figure. Despite this, pulse
 12 has disappeared entirely by $2.5$ $\times$ $10^{4}$ seconds, whilst the
 earlier pulse can still be seen on the axes, even though the modelled
 emission is at a level significantly below the total SED. The bottom two
 panels show late times in the light curve, where only the afterglow
 contributes flux to the total SED. Looking only at the afterglow in all six
 panels, it can be seen that it varies very little at early times. This is due
 to the combination of a quick rise time after an early launch with a late
 transition time between the exponential and power-law decay phase for the
 afterglow component. This implies that there is a period where the afterglow
 effectively plateaus for a few thousand seconds. This behaviour can be
 identified in the light curves shown in Figure \ref{fit_ag}.\par
The SEDs in Figure \ref{sed1} suggest that the optical emission is
 fitted by the extrapolation of the afterglow Band function, particularly at
 late times. The second panel is perhaps the most puzzling as the optical
 and near infrared data are aligned approximately with the afterglow SED,
 however there is a clear dip in the light curves, which suggests that a
 component which simply rises then falls smoothly cannot explain the morphology
 that is seen. We have not tried to fit a low energy break to the afterglow
 Band function, as discussed previously, because such a break is expected at
 energies well below the $H$-band, due to the optically thin nature of the
 circumburst medium. Additionally, the correct level of flux has been
 reproduced without the introduction of a further spectral break.\par

\subsection{Early time prompt dominated fit}

By fitting the optical and IR emission with only a significant afterglow
 component, there were discrepancies between the model and the
 data. Namely the potential decay before the sharp rise
 observed in the $V$-band and the deficit in flux observed in the $I$-, $V$-
 and $R$-bands between 300 and 2,000 seconds. In addition to this, the
 $U$-band data are over predicted, which is possibly due to the absorption
 correction from the Lyman forest as previously discussed, or alternatively
 could be better described with an alternative model. In an attempt to
 explain these features, we considered a model where these early-times were
 dominated by the prompt emission in the optical and near infrared regimes.\par

\subsubsection{Light Curves}

We first only considered the prompt pulses, by turning off the afterglow
 component and fitting only prior to 2,000 seconds. Once the early time
 emission was fitted, we then included the afterglow to account for the
 late time emission. The potential for degeneracy between $p_{a}$ and
 $E_{a}$ was considered, as the energy difference between the X-ray data and
 the optical and IR bands is over two orders of magnitude, while the $U$- and
 $K$-bands are only separated by less than a single order of magnitude. Data
 were not taken simultaneously in these two bands which are the most
 spectrally separated. It is therefore difficult to
 determine both $E_{a}$ and $p_{a}$ independently. For this reason the
 break energy was set to a value of 0.3 $\keV$ (at $T_{pk}$) which is the
 soft end of the XRT spectral range. This implies that the fit to the
 high-energy regime remains unaltered, whilst giving the maximum energy range
 in the SED between $E_{a}$ and the optical and IR bands. Therefore the value
 for $p_{a}$ obtained is a lower limit, as moving $E_{a}$ to lower values
 would steepen the power-law index in the low energy part of the
 spectrum.\par
\begin{figure*}
\begin{center}
\includegraphics[height=15cm,angle=0]{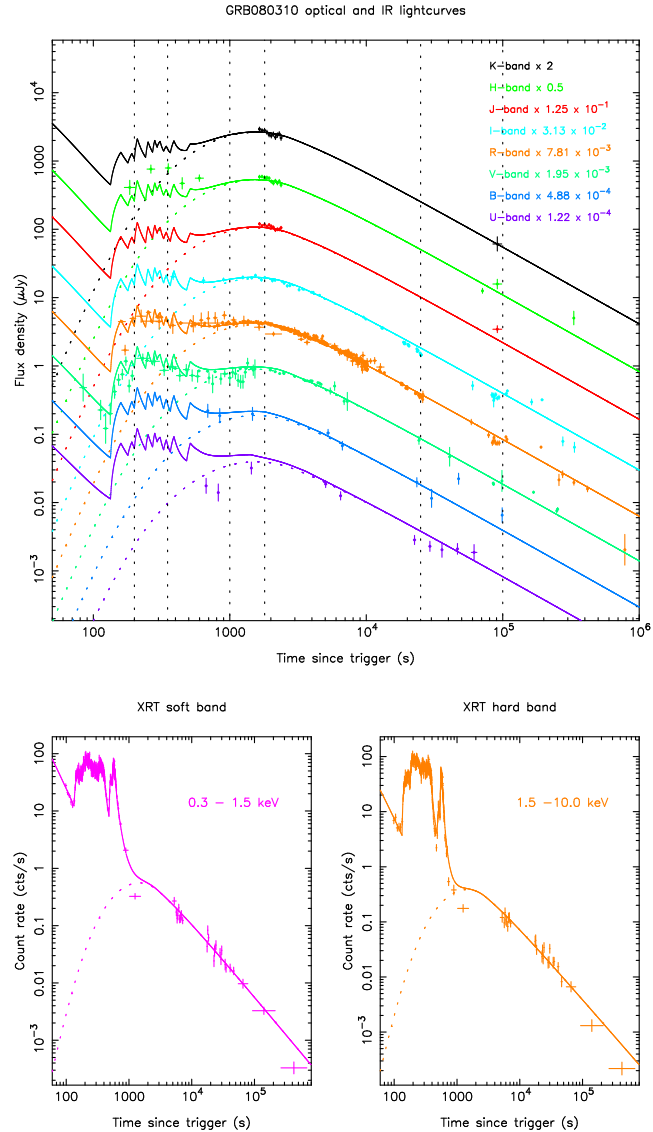}
\end{center}
\caption{Simultaneous fit to optical, IR and XRT data, in which the
 early time optical and IR are fitted by the prompt emission. Top panel:
 Optical and IR bands fitted with low energy break ($K$, $H$, $J$, $I$, $R$,
 $V$, $B$ and $U$ bands). Bottom left: XRT soft band fit. Bottom right: XRT
 hard band fit. $E_{a}=$ 0.3 keV, $p_{a}$ and $d_{a}$ are allowed to float
 (producing a value of $d_{a}=$ -1.01) and the afterglow parameters ($T_{r}$,
 $T_{a}$, $\alpha_{a}$ and $F_{a}$) are also fitted.
 The reduced $\chi^{2}$ for this fit is 2.43 with 1,642 degrees of freedom.}
\label{fit_prompt}
\end{figure*}

Initial attempts to model the low energy flux with prompt pulses held the
 parameters $p_{a}$, $d_{a}$ and $E_{a}$ to the same value for all pulses.
 This led to a bulk value of $p_{a}$ = $0.65_{-0.02}^{+0.02}$, which
 is significantly different from the spectral index of the Band function below
 $E_{pk}$, but not consistent with the expected values of $p_{a}$ if
 self-absorption was having an effect on the emission that is observed. The
 1$\sigma$ limits on this value do not include $\frac{1}{3}$, which is what
 would be expected if there is an absence of self-absorption, and the optical
 and IR bands are below the emission frequency of electrons with the minimum
 energy in the electron population. The reduced $\chi^{2}$ value is 3.61 in
 this instance, with 1,642 degrees of freedom. There were still some issues
 with this fit, such as the fit not describing the slight decay followed by
 sharp rise in flux seen in the $V$-band. So we decided to allow all pulses to
 have independent values of $p_{a}$, to see if this could account for the
 features seen in the data. The method adopted to do this remained the same as
 that for the bulk value of $p_{a}$, and the light curves obtained are shown in
 Figure \ref{fit_prompt}.\par
Given the choice of two afterglow models; the original functional form of
 \citet{2007ApJ...662.1093W} and the modified version shown in Eqn
 \ref{ageqn}, we found that in this instance, the better fit was obtained
 with the former, as the smoother peak to the afterglow component allowed
 for a closer fit to some of the $J$-, $K$- and $H$-band data at this time.
 This also removed a free parameter from the fitting routine. We still
 retained the difference between optical and high energy $\alpha_{a}$ to
 describe the late time decay. With these modifications, the early time prompt
 dominated model had a reduced $\chi^{2}$ of 2.43 for 1,642 degrees of
 freedom. This is significantly better than the prompt model with a bulk
 value of $p_{a}$ for all pulses, with a change in total $\chi^{2}$ of over
 1,850.\par
There are little data to constrain $p_{a}$ for the first four pulses,
 which are predominantly observed with BAT, and so these values have
 large associated uncertainties. Despite this, it was clear that a large value
 of $p_{a}$ was required for pulses 1, 2 and 4, to allow for the
 decay and subsequent rise seen in the $V$-band data. The best fit to the
 initial decay was found by modelling it with pulse two, and suppressing
 pulses one and three so that they did not contribute to the optical and IR
 emission observed from the beginning of the optical coverage. When actually
 fitted, the values of $p_{a}$ for pulses 1, 2 and 4 all were larger than
 $\frac{5}{2}$, and so in the same manner as all the prompt pulses in the
 afterglow-dominated model, these were set to $p_{a}$ = $\frac{5}{2}$ as Figure
 \ref{fit_prompt} shows that the sharp rise is now well fitted by the
 rise of pulse four in a similar way to the simultaneous rise seen in
 the XRT data. The data obtained from KAIT have an earlier point at 57 seconds,
 which does not show this decay, it is difficult to fit this individual point,
 as whilst it is at an ideal time to investigate the prompt optical behaviour
 of the GRB, it is a 30 second observation, and so it is an average over a time
 during which large variations might be expected. This is why we have not tried
 to fully explain this point with our model and have removed it from the
 fitting undertaken. Indeed, non-detections from both
 Super-LOTIS and RAPTOR \citep{2008GCN..7403....1W} at similar times to this
 KAIT datapoint offer further evidence towards the faintness of the emission
 in the optical regime at this time.\par
Analysis of the residuals for the optical, IR, XRT and BAT data show there
 are three clear contributions to the total $\chi^{2}$ statistic. The first
 contribution to the fit statistic is from the XRT afterglow component of the
 model. This deviates from the observed data at two times. Firstly, the
 transition from the rapid decay phase to the afterglow shows an over
 prediction by the
 model, as do the data after $10^{5}$ seconds. The latter of the two features
 could be explained by a late temporal break in the X-ray afterglow. Such a
 break should be achromatic, and therefore also observed in the late-time
 optical data. The final $R$-band observation, at approximately nine days
 after the trigger, is fainter than the extrapolation from the final power-law
 decay in the afterglow-dominated model, which would support the possibility of
 a jet break. With the prompt-dominated optical and IR model, there is a
 suggestion of either a plateau or rebrightening at approximately one day
 after the trigger, followed by a steeper decay. In both models, the data are
 too sparse to justify the inclusion of further parameters in the late-time
 modelling of the optical data.\par
The next significant contribution to the value of $\chi^{2}$ comes from the
 early-time REM data points between 150 and 600 seconds in the $H$-band. These
 are under predicted by the model as shown in Figure \ref{fit_prompt}. Due to
 the sparse nature of the NIR data at this time, the fit is driven by the more
 densely sampled $V$- and $R$-band data. This discrepancy is better
 highlighted in the SEDs shown in Figure \ref{sed2}.\par
The other notable source of contribution to the fit statistic is
 the KAIT data. The data were initially reported in unfiltered magnitudes,
 with values ranging between 17$^{m}$ and 20$^{m}$, and errors typically less
 than a tenth of a magnitude. With such small errors for over two hundred data
 points the contribution to the total validity of the fit from the intrinsic
 scatter of the dataset is significant, despite the fact that the data and
 model can be seen to agree by eye.\par

\subsubsection{Spectral energy distributions}

\begin{figure*}
\begin{center}
\includegraphics[height=20.0cm,angle=0]{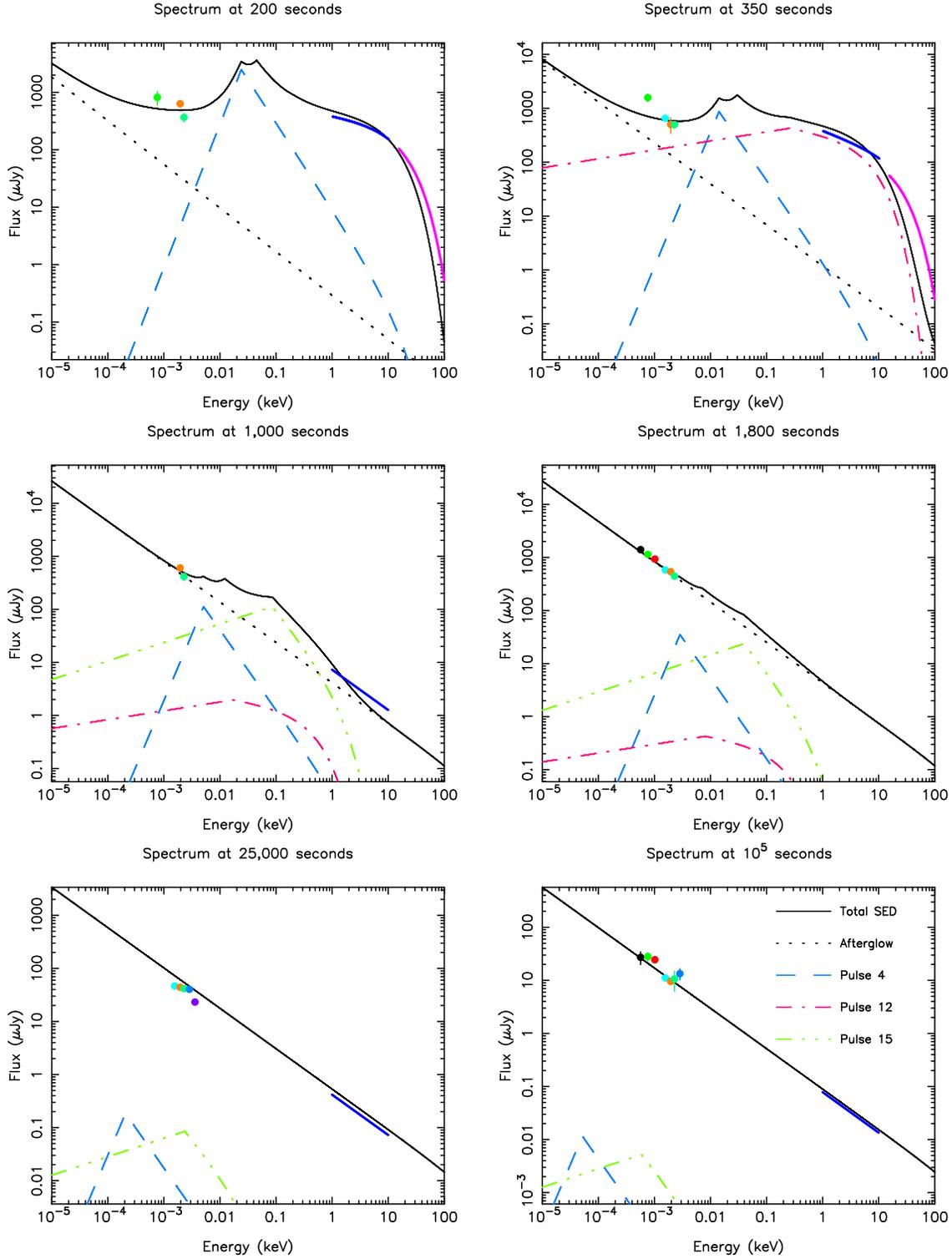}
\end{center}
\caption{SEDs for fiducial times in the light curve of GRB~080310 using the
 prompt-dominated early time model. The solid line is the total SED and the
 dotted line is the contribution from the afterglow component. Pulses 4, 12 and
 15 are also shown by the dashed and dotted lines as outlined in the key shown
 in the bottom right panel. These are included as examples to show the
 evolution of some of the longer lasting pulses in the fit. The colours of
 the data points follow the same scheme as that used in the light curves, with
 the BAT and XRT data being represented by lines which extend along their
 entire coverage.}
\label{sed2}
\end{figure*}

As can be seen in the SEDs shown in Fig. \ref{sed2}, by introducing the prompt
 component to the spectrum of GRB~080310, each SED has structure beyond the
 simple Band function of the afterglow at earlier times in the low energy
 bands. For the prompt-dominated model of the early-time low energy emission,
 the spectral indices typically lie in the range -0.65
 $<$ $p_{a}$ $<$ 1. This does not include the earliest pulses, which have a
 necessarily steep spectral index to prevent their emission dominating (and
 poorly fitting) the optical and IR light curves. These have been fixed to a
 value of $p_{a}$ = $\frac{5}{2}$, as well as pulses 8 and 16, which also
 require steep indices.\par
Unlike in the afterglow-dominated model, the SED at 350 seconds shows the
 optical and IR (particularly the $H$-band) model to disagree with the data;
 this is the most valuable of the SEDs in Figure \ref{sed2}, as it is one of
 the only two prior to the afterglow making the dominant contribution to the
 total low energy emission, and of these two has the best low energy coverage.
 The $R$-, $I$- and $V$-band data at this time are at approximately the correct
 level of flux for each band, although the $I$-band and $R$-band deviate from
 the expected shape of the SED. However the $H$-band data (which offers the
 largest range of spectral information when considered with the other three
 bands) is noticeably under predicted, as corroborated by the light curves
 in Figure \ref{fit_prompt}. When inspected by eye, the four bands at this time
 could be thought to lie on a single power-law with a spectral index similar to
 that of the afterglow. This would lend more credence to the afterglow
-dominated early-time optical and near infrared model.\par
The subsequent SEDs are all dominated by the Band function of the afterglow at
 optical and IR energies, and, despite the use of the original 
\citet{2007ApJ...662.1093W} afterglow model and changes in $T_{a}$, $F_{a}$,
 and $T_{r}$, the late time fit is similar to that obtained for the afterglow
-dominated fit. The optical and near infrared data points are all over
 predicted at $2.5$ $\times$ $10^{5}$ seconds, which something that can be
 seen by looking at the light curves, particularly in the case of the UVOT
 $u$-band data. The $u$-band dataset displays some unusual characteristics,
 being significantly fainter than expected prior to 1,000 seconds and then
 appearing to plateau between $2$ $\times$ $10^{4}$ and $1$ $\times$ $10^{5}$
 seconds. We considered the removal of this dataset from the analysis, but
 as both the values of parameters obtained for the best fit seem largely
 insensitive to the $u$-band data we retained them. This is a result of the
 relatively sparse coverage in the $u$-band, principally in comparison to
 that of the $R$-band, the weighting of the $u$-band data in determining the
 fit statistic is small.\par

\section{Discussion}

For direct comparison between the two models discussed in the previous section,
 their parameters have been included in Table \ref{tab2}. Given the
 degeneracy between $E_{a}$ and $p_{a}$ it isn't possible to simultaneously
 fit exact values to both without a more extensive low energy coverage over a
 larger range of wavelengths. Because of this, $E_{a}$ was fixed
 at 0.3 keV, in an attempt to find a value of $p_{a}$.\par
 \begin{table*}
 \centering
  \caption{Pulse $p_{a}$ values and afterglow parameters for both the prompt
 and afterglow dominated early time optical emission models. Note that the
 value for $T_{l}$ in the prompt dominated model is untabulated, as the
 \citet{2007ApJ...662.1093W} model was used for the afterglow in this
 instance. All instances where $p_{a}$ = $\frac{5}{2}$ have the values fixed
 rather than fitted, and so no uncertainties were calculated for them.}
  \label{tab2}
  \begin{tabular}{@{}cccc}
    \hline
    Parameter & Parameter description & Afterglow & Prompt \\
     & & dominated & dominated \\
    \hline
    $p_{a}$ (pulse 1) & X-ray to optical spectral index & 2.5 & 2.5 \\
    $p_{a}$ (pulse 2) & X-ray to optical spectral index & 2.5 & 2.5 \\
    $p_{a}$ (pulse 3) & X-ray to optical spectral index & 2.5 &
 -0.65$^{+0.59}_{-0.59}$ \\
    $p_{a}$ (pulse 4) & X-ray to optical spectral index & 2.5 & 2.5 \\
    $p_{a}$ (pulse 5) & X-ray to optical spectral index & 2.5 &
 0.03$^{+0.04}_{-0.07}$ \\
    $p_{a}$ (pulse 6) & X-ray to optical spectral index & 2.5 &
 0.11$^{+0.46}_{-0.17}$ \\
    $p_{a}$ (pulse 7) & X-ray to optical spectral index & 2.5 &
 -0.11$^{+0.43}_{-0.25}$ \\
    $p_{a}$ (pulse 8) & X-ray to optical spectral index & 2.5 &
 2.5 \\
    $p_{a}$ (pulse 9) & X-ray to optical spectral index & 2.5 &
 -0.06$^{+0.09}_{-0.01}$ \\
    $p_{a}$ (pulse 10) & X-ray to optical spectral index & 2.5 &
 -0.01$^{+0.26}_{-0.26}$ \\
    $p_{a}$ (pulse 11) & X-ray to optical spectral index & 2.5 &
 -0.11$^{+1.00}_{-0.30}$ \\
    $p_{a}$ (pulse 12) & X-ray to optical spectral index & 2.5 &
 0.17$^{+0.19}_{-0.11}$ \\
    $p_{a}$ (pulse 13) & X-ray to optical spectral index & 2.5 &
 1.0$^{+0.22}_{-0.22}$ \\
    $p_{a}$ (pulse 14) & X-ray to optical spectral index & 2.5 &
 0.04$^{+0.11}_{-0.08}$ \\
    $p_{a}$ (pulse 15) & X-ray to optical spectral index & 2.5 &
 0.35$^{+0.08}_{-0.09}$ \\
    $p_{a}$ (pulse 16) & X-ray to optical spectral index & 2.5 &
 2.5$^{+}_{-}$ \\
    $T_{l}$ (seconds) & Afterglow launch time & 120$^{+9}_{-11}$ & - \\
    $T_{r}$ (seconds) & Afterglow rise time & 223$^{+16}_{-20}$ &
 1000$^{+435}_{-131}$ \\
    $T_{a}$ (seconds) & End time of afterglow plateau phase &
 6368$^{+501}_{-495}$ & 2937$^{+1056}_{-1346}$ \\
    $\alpha_{a}$ & Afterglow temporal decay index & 1.30$^{+0.01}_{-0.01}$ &
 1.31$^{+0.01}_{-0.01}$ \\
    $\Delta\alpha_{a}$ & $\Delta\alpha_{a}$ between optical and X-ray decays &
 0.21$^{+0.03}_{-0.04}$ & 0.18$^{+0.1}_{-0.03}$ \\
    $F_{a}$ ($\gamma$.cm$^{-2}$.s$^{-1}$) & Integrated Flux at $T_{a}$
 & 2.34$^{+0.08}_{-0.08}$ $\times$ 10$^{-2}$ & 2.77$^{+0.23}_{-0.16}$
 $\times$ 10$^{-2}$ \\
    \hline
\end{tabular}
\end{table*}

There is no clear difference in the fit statistic for either of the models
 presented in this work. The difference in total $\chi^{2}$ is less than 30
 over such a large range of degrees of freedom. Whilst numerically, this
 difference is in favour of the prompt-dominated early-time optical and
 near infrared model, we do not believe that the magnitude of the difference
 is enough to warrant favouring one model above the other.\par
For the afterglow-dominated early-time model the fitted bulk low-energy
 spectral index for the prompt pulses is unphysically steep ($p_{a}$ = 5.6).
 However, the change in fit statistic when forcing this value to one consistent
 with self-absorption was near negligible, which is why we therefore adopted
 the less extreme value. In contrast, most values of $p_{a}$ are more
 reasonable (-0.65 $<$ $p_{a}$ $<$ 1) when the early-time low-energy emission
 is dominated by the prompt pulses identified at higher energies. This is not,
 however, true for all the pulses. It was necessary to suppress pulses 1, 2, 4,
 8 and 16 with positive, steep low energy spectral indices so that the optical
 and IR emission was not over predicted. To understand this behaviour we
 looked at the one dimensional $\chi^{2}$ distribution for $p_{a}$ of the
 earliest of these three pulses. We found that the distribution
 reduced asymptotically at higher values of $p_{a}$, meaning that there was
 no clear minimum. However, the actual reduction in the fit statistic achieved
 by reducing the low energy spectral index rapidly dropped, so this decrease
 in $\chi^{2}$ value quickly stopped making a significant difference to the
 quality of the total fit compared to other contributions to the $\chi^{2}$
 statistic. After establishing this, we again adopted values of $p_{a}$ =
 $\frac{5}{2}$ in these instances, with a minimal increase of $\chi^{2}$ being
 introduced as a result.\par
Given the variations in $p_{a}$ for the prompt dominated model, we tried to
 relate these values to other parameters of the pulses, such as pulse duration
 ($T_{pk}$) and the low energy photon index of the Band functions for each
 pulse ($b_{1}$). We found no correlations between $p_{a}$ and the other
 pulse parameters.\par
\begin{figure}
\begin{center}
\includegraphics[height=8cm,angle=270]{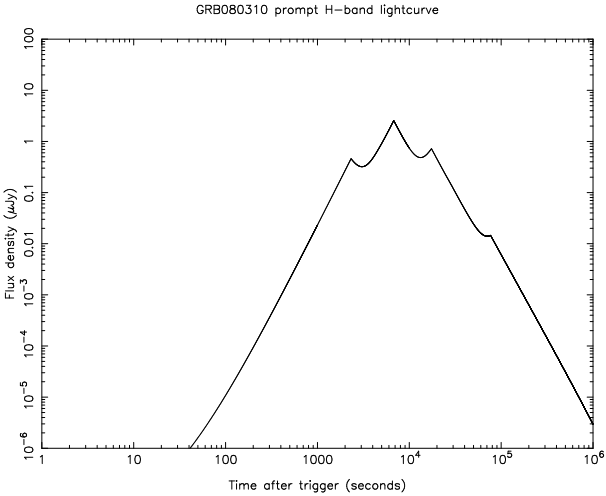}
\end{center}
\caption{Modelled $H$-band light curve showing the prompt component only for the
 afterglow dominated fit.}
\label{promptopt}
\end{figure}

In the afterglow dominated fit, the prompt component is several orders of
 magnitude fainter than the observed emission. Additionally, with positive
 values of
 $p_{a}$, time is required for the peak of the pulse spectrum to migrate to
 the optical and IR bands, before which the spectrum is rising at these
 energies. The pulse emission is largest around $E_{a}$, the evolution of
 which is governed by the temporal index $d_{a}$, which has been shown to be
 negative. This means that the pulse energy decreases with increasing time. In
 addition to this, the normalization of each pulse scales with $t^{-1}$, which
 means that when the peak of emission reaches the optical part of the spectrum
 the total flux is reduced. To illustrate the prompt component in this
 scenario, we subtracted the afterglow component from the afterglow-dominated
 model, and produced Figure \ref{promptopt}, which shows the $H$-band
 prompt-only light curve that underlies the afterglow emission.\par
 The afterglow-dominated, early-time optical model required three
 alterations to the afterglow component; a change of temporal power-law index
 between high and low energies, a power-law rise at early times and a variable
 launch time $T_{l}$. However, even with these modifications, there were
 several features of the data that were not entirely modelled in this scenario,
 including the apparent decay in the $V$-band prior to the rapid rise and the
 slight deficit in flux between 300 and 2,000 seconds. The tendency towards
 unphysical values of $p_{a}$ also suggested that fitting the early-time
 emission with a fast rising afterglow may not provide the best model of the
 observed flux. The SEDs of Figure \ref{sed1} confirm that the
 afterglow-dominated fit provides a reasonable fit to the data and
 that at 350 seconds, which is a time where at high energies the prompt
 components are dominant, the afterglow Band function fits both the level and
 slope of the SED.\par
When fitting the early-time flux with the prompt pulse emission model we first
 adopted a single value of $p_{a}$ which was held constant for all pulses. The
 low energy spectral index obtained was $p_{a}=$0.65, however most of the
 early time emission was hidden by the tail of one of the earliest prompt
 pulses. Aside from a poor fit to the $H$-band data, the fast rise seen in the
 $V$-band was not well represented in this instance, so the pulses were allowed
 to have independent values of $p_{a}$. This produced a better fit to the data,
 significantly improving the $\chi^{2}$ statistic. The fast rise in the
 $V$-band, and the small scale variations in the light curves seem to be
 represented well by this model. There were, however, discrepancies with this
 alternative. Firstly, the $H$-band data were under predicted by the model,
 which can be seen in both the light curves and SEDs shown for this model.
 The SEDs also show that, at 350 seconds, the optical SED appears to be more
 accurately represented by a single power-law which agrees more with the
 nature of the afterglow. Three of the first four pulses (and also pulses 8 and
 16) tend towards an unphysically steep spectral index when fitted. As with the
 afterglow-dominated model, constraining the values of $p_{a}$ to $\frac{5}{2}$
 in these instances does not significantly reduce the quality of the fit
 obtained.\par
The other values of $p_{a}$ for the prompt-dominated fit could be consistent
 with the energy bands lying below the energy of emission from the least
 energetic electron in the relativistic population responsible for the
 observed photons. We found no relation between the low energy spectral index
 for each pulse and any other pulse parameter, giving no insight into the
 cause of the variation of $p_{a}$. An interesting result is the marked
 difference between pulses 1, 2, 4, 8 and 16 when compared with all the other
 pulses. $p_{a}$ for these early pulses indicates their emission is from an
 optically thick environment in the optical regime. Another implication is
 that these pulses must be launched from a different environment than the
 other eleven which have far shallower spectral indices at low energies.\par
One additional benefit to the prompt-dominated model was the ability to
 return to the simpler afterglow model of \citet{2007ApJ...662.1093W}, as the
 rapid rise seen in the $V$-band was attributed to the rise of pulse four.\par
An alteration attempted with the prompt-dominated early-time model was to
 introduce an offset to the onset of the afterglow. It was hoped that by
 allowing the afterglow to rise earlier, the SED at 350 seconds could
 reconciled with the model. This was found not to be possible, as this
 returned the model to something resembling the afterglow-dominated fit, and
 therefore we were unable to model the small scale variations and observed
 optical deficit in the plateau between 300 and 2,000 seconds.\par
Both of the alternative models presented in Figures \ref{fit_ag} to
 \ref{sed2} are poor at fitting the transition between the rapid decay phase
 and the afterglow plateau observed in the X-rays. We attempted to better
 model this by allowing the afterglow to rise at a later time than presented
 in either of the two fits highlighted in Table \ref{tab2}. By doing so, it
 was possible to improve the quality of the high energy afterglow fit, but to
 the detriment of that obtained for the optical data. With an afterglow that
 rises later but more quickly, the optical data between 3,000 seconds and
 8,000 seconds are under predicted by the model suggested. With a larger
 number of data points, and correspondingly better statistics, the optical
 points were those that we therefore favoured. The prompt-dominated,
 early-time optical and IR model described in Table \ref{tab2} and illustrated
 in Figures \ref{fit_prompt} and \ref{sed2} summarize this model.
 Unfortunately, the XRT coverage contains a gap during the afterglow plateau
 phase, and therefore the exact morphology of this component, which would
 help significantly constrain the rise of the afterglow, is unknown at these
 times.\par
An alternative suggestion, given the acceptable fit at approximately 2,000
 seconds in all the available optical IR bands, to explain the factor of two
 or three difference between the X-ray data and model at this time would be
 to include a spectral break in the afterglow spectrum. Whilst this may
 improve the fit at these times, the later afterglow coverage between
 10$^{4}$ and 10$^{5}$ seconds show both the higher and lower energy bands to
 be modelled at the correct level.\par
Additionally, a large contribution to the $\chi^{2}$ distribution was from the
 optical and IR data, particularly the KAIT dataset, with low magnitude errors.
 Whilst the observational errors in these datapoints may be as reported, each
 of the datasets had to be calibrated so they all were in standard bands. In
 doing so the systematic errors associated with these data points increased.
 We considered this, and added a systematic of 0.03$^{m}$ to every optical or
 IR point. The quoted values for the fit statistics include this systematic
 source of error.\par
Knowing the characteristic times of the afterglow for each of the two models
 allows us to calculate the initial bulk Lorentz factor in both an interstellar
 medium (ISM) like or wind dominated circumburst environment. To do so we
 used Eqns \ref{eqn_ism} \citep{2007A&A...469L..13M} and \ref{eqn_wind}
 \citep{1999ApJ...520..641S}.\par
\begin{equation}
  \frac{\Gamma_{0,ISM}}{2}=\left(\frac{3E_{iso}\left(1+z\right)^{3}}
{32\pi nm_{p}c^{5}\eta t_{peak}^{3}}\right)^{\frac{1}{8}},
\label{eqn_ism}
\end{equation}
\begin{equation}
  \frac{\Gamma_{0,wind}}{2}=\left(\frac{E_{iso}\left(1+z\right)}
{8\pi Am_{p}c^{3}\eta t_{peak}}\right)^{\frac{1}{4}}.
\label{eqn_wind}
\end{equation}
In the equations above $E_{iso}$ is the isotropic equivalent energy of the GRB,
 $\eta$ is the radiative efficiency of the fireball,
 $z$ is the measured redshift, $n$ is the number density of the circumburst
 medium and $A$ is a normalization for the density in the wind-like case (where
 $\rho~\propto~r^{-2}$). As with \citep{2007A&A...469L..13M}, we assume that
 $n$ = 1 cm$^{-3}$, $\eta$ = 0.2 and $A$ = 3 $\times$ 10$^{35}$ cm$^{-1}$.\par
$t_{peak}$ is the time at which the afterglow peaks, and can be calculated
 using Eqn \ref{eqn_tpk}, taken from \citet{2007ApJ...662.1093W}:\par
\begin{equation}
t_{peak}=\left(\frac{T_{r}T_{a}}{\alpha_{a}}\right)^{\frac{1}{2}}.
\label{eqn_tpk}
\end{equation}
The derived values of $t_{peak}$ and the initial bulk Lorentz factors are
 shown in Table \ref{tab_lor}.
 \begin{table}
 \centering
  \caption{$t_{peak}$ values and initial bulk Lorentz factors for both the
 prompt and afterglow-dominated early-time optical emission models in both
 an ISM and wind like environment.}
  \label{tab_lor}
  \begin{tabular}{@{}cccc}
    \hline
    Model & $t_{peak}$ (s) & $\Gamma_{0,ISM}$ & $\Gamma_{0,wind}$ \\
    \hline
    Prompt & $1497^{+845}_{-714}$ & $169.1^{+41.5}_{-21.0}$ &
 $75.1^{+34.6}_{-11.1}$ \\
    Afterglow & $1044.^{+112}_{-122}$ & $193.6^{+43.3}_{-6.1}$ &
 $82.2^{+36.6}_{-4.1}$ \\
    \hline
  \end{tabular}
 \end{table}
The calculated values show that an ISM type circumburst medium leads to higher
 Lorentz factors, which is consistent with the results shown for a larger
 sample by \citet[Submitted]{philsub}. By definition the
 afterglow-dominated model peaks at an earlier time, which means that the
 initial bulk Lorentz factor is necessarily higher for this case, as
 demonstrated in Table \ref{tab_lor}. It is worth noting, however, that when
 considering the errors quoted, the prompt-dominated values cannot be said to
 be distinct from the corresponding values derived from the afterglow-dominated
 fit.\par

\section{Conclusions}

In this work, we have taken the pulse model of \citet{2009MNRAS.399.1328G} and
 used it in a similar manner to \citet{2010MNRAS.403.1296W} to reproduce the
 prompt light curve of GRB~080310. Combining it with an afterglow model
 \citep{2007ApJ...662.1093W}, we have tried to produce a simultaneous fit to
 not only the data from the \swift\ XRT and BAT instruments, but also the
 available optical and IR datasets. The aims behind this work were to
 establish the origin of the observed early-time optical and IR emission, and
 to attribute it to either the prompt or afterglow component of the GRB.\par
The first conclusion of this work is that a low-energy break is required in the
 spectra of prompt pulses in order to fit the optical and near IR flux,
 regardless of the origin of this emission.\par
The simplest model considered for the optical and IR was to use this low
 energy break to remove the prompt component entirely from the observed
 emission. Whilst this successfully recreated the broad scale structure of the
 optical and IR light curves, and the SEDs also appear satisfactory, the
 value of $p_{a}$ to which the fitting tended to was unphyiscal ($p_{a}$ = 5.6)
 which is inconsistent with that expected for a realistic spectrum, such as
 synchrotron radiation below the minimum emitted photon energy ($E_{m}$) for
 which $p_{a}$ should be $\frac{1}{3}$ or synchrotron self absorption, where
 $p_{a}$ is expected to be $\frac{5}{2}$ or 2, when the self absorption
 frequency is above or less than $E_{m}$ respectively. However, having
 looked at the one dimensional $\chi^{2}$ surface for $p_{a}$, we found
 that the distribution asymptotes to a better fit at large values of $p_{a}$.
 As a result we used a more realistic value of $p_{a}$ = $\frac{5}{2}$, which
 didn't significantly reduce the quality of the fit. The implications of this
 are that self absorption is a necessary mechanism in order to fit the
 optical and near infrared flux observed, when assuming the prompt pulses of
 the high energy light curves do not contribute at early times in the lower
 energy bands.\par
 Morphological inconsistencies in the light curves required the exploration of
 an alternative solution. This alternative was to allow the prompt emission to
 dominate the early times of the optical and IR light curves. An initial
 treatment of the prompt radiation, in which a single value for $p_{a}$ was
 assigned to all pulses, proved insufficient to fit the data. Following this,
 by allowing the pulses to have independent values of $p_{a}$, a fit
 of similar statistical merit to the afterglow-dominated model was obtained
 (Figures \ref{fit_prompt} and \ref{sed2}). With this model, an additional
 break was still required for all the prompt pulses, but with a variety of
 values of $p_{a}$. For five pulses (particularly three of the earliest four)
 steep spectral indices were required, tending to unphysical values when
 fitted. As before, in these instances, we found in these instances that it
 was possible to fix the appropriate values of $p_{a}$ to $\frac{5}{2}$ without
 significantly altering the quality of the fit. This again suggests that self
 absorption could be an important mechanism by which the prompt emission is
 suppressed in the optical and infrared regimes. For those pulses whose
 spectrum required a break, but not to the extent of pulses 1, 2, 4, 8 or 16,
 we suggest that the break energy is at a value between the optical and
 X-ray bands, but given the degeneracy between $p_{a}$ and $E_{a}$ have not
 fitted it. These pulses could then also have a value of $p_{a}$ =
 $\frac{5}{2}$, but peak at an energy nearer to that of the optical bands.\par
From the results obtained, it is unclear whether the optical and IR emission
 of GRB~080310 originates from central engine or afterglow activity.
 Neither case accurately describes all of the data. The afterglow-dominated
 model is insufficient to describe all of the structure seen in the optical
 light curves, however, the SEDs produced, particularly at 350 seconds suggest
 that the optical and near infrared are more faithfully represented by
 afterglow emission. To help discriminate between prompt and
 afterglow emission as a source for early time emission a similar analysis is
 required on a larger sample of GRBs. Bursts  in such a sample have several
 important pre-requisites. Firstly, good continuous optical data are required
 from very early to late times, preferably in several bands, simultaneously.
 Having such a data point from KAIT though, highlights that at the earliest
 times high temporal resolution is required too, as GRBs are highly variable
 during their prompt phases.\par
GRBs which will be the best candidates for further analysis will be those that
 contain pulse structure in the optical light curves. If these pulses are
 simultaneous with similar structure at higher energies, then it is likely
 they share a common origin. In contrast to this, should the pulses occur at
 markedly different times then an alternative mechanism must be found to
 explain their behaviour. The ideal type of burst for this analysis would
 therefore be one which has a long duration as observed in BAT (or
 \textit{Fermi} GBM) and exhibits strong flaring behaviour after the first
 hundred seconds in the X-ray regime. Such times are not only feasible for
 ground based follow-up, but also allow for sufficient temporal resolution to
 discern any features at these lower energies.\par

\section*{Acknowledgements}

We would like to thank the referee for their useful comments. This work is
 supported at the University of Leicester by the STFC.
\begin{bibliography}{080310}
  \bibliographystyle{mn2e}
\end{bibliography}
\appendix
\section{Optical and near infrared data }
Data from the \swift\ satellite are publically available. In addition to this
 Tables \ref{opt_k} - \ref{opt_u} detail the ground-based observations taken
 in the optical and near IR regimes, alongside the \textit{Swift}-UVOT
 observations used in this analysis.\par
\onecolumn
\begin{table*}
  \centering
    \caption{All near infrared observations which were calibrated to the
      $K$-band, showing the central time after trigger of each exposure,
      exposure time ($T_{exp}$), filter,
      instrument, flux and flux errors. The filters quoted are the original
      filters that observations were taken, before conversion to the standard
      filters used in the later analysis. Fluxes are extinction corrected and
      any external references are cited. A full version of this table is
      available online.}
    \label{opt_k}
    \begin{tabular}{@{}ccccccc}
      \hline
      Time after trigger (s) & $T_{exp}$ (s) & Filter & Instrument & Flux ($\mu$Jy) & Flux error ($\pm\mu$Jy) & External sources \\
      \hline
      1650 & 47 & $K$ & PAIRITEL & 1469.20 & 101.21 & \\
      1723 & 47 & $K$ & PAIRITEL & 1395.98 & 95.66 & \\
      1796 & 47 & $K$ & PAIRITEL & 1405.27 & 103.07 & \\
      1868 & 47 & $K$ & PAIRITEL & 1315.23 & 96.35 & \\
      1941 & 47 & $K$ & PAIRITEL & 1212.27 & 92.56 & \\
      2013 & 47 & $K$ & PAIRITEL & 1171.55 & 92.63 & \\
      \hline
    \end{tabular}
\end{table*}
\begin{table*}
  \centering
    \caption{All near infrared observations which were calibrated to the
      $H$-band, showing the central time after trigger of each exposure,
      exposure time ($T_{exp}$), filter,
      instrument, flux and flux errors. The filters quoted are the original
      filters that observations were taken, before conversion to the standard
      filters used in the later analysis. Fluxes are extinction corrected and
      any external references are cited. A full version of this table is
      available online.}
    \label{opt_h}
    \begin{tabular}{@{}ccccccc}
      \hline
      Time after trigger (s) & $T_{exp}$ (s) & Filter & Instrument & Flux ($\mu$Jy) & Flux error ($\pm\mu$Jy) & External sources \\
      \hline
      185 & 36 & $H$ & REM-REMIR & 820.92 & 222.12 & \\
      264 & 36 & $H$ & REM-REMIR & 1521.60 & 225.43 & \\
      354 & 36 & $H$ & REM-REMIR & 1578.70 & 250.66 & \\
      448 & 36 & $H$ & REM-REMIR & 942.54 & 201.13 & \\
      598 & 86 & $H$ & REM-REMIR & 1122.79 & 131.21 & \\
      1650 & 47 & $H$ & PAIRITEL & 1174.00 & 65.33 & \\
      \hline
    \end{tabular}
\end{table*}
\begin{table*}
  \centering
    \caption{All near infrared observations which were calibrated to the
      $J$-band, showing the central time after trigger of each exposure,
      exposure time ($T_{exp}$), filter,
      instrument, flux and flux errors. The filters quoted are the original
      filters that observations were taken, before conversion to the standard
      filters used in the later analysis. Fluxes are extinction corrected and
      any external references are cited. A full version of this table is
      available online.}
    \label{opt_j}
    \begin{tabular}{@{}ccccccc}
      \hline
      Time after trigger (s) & $T_{exp}$ (s) & Filter & Instrument & Flux ($\mu$Jy) & Flux error ($\pm\mu$Jy) & External sources \\
      \hline
      1650 & 47 & $J$ & PAIRITEL & 961.77 & 44.42 & \\
      1723 & 47 & $J$ & PAIRITEL & 903.55 & 41.16 & \\
      1796 & 47 & $J$ & PAIRITEL & 932.98 & 47.86 & \\
      1868 & 47 & $J$ & PAIRITEL & 879.73 & 43.09 & \\
      1941 & 47 & $J$ & PAIRITEL & 928.43 & 44.19 & \\
      2013 & 47 & $J$ & PAIRITEL & 887.46 & 43.31 & \\
      \hline
    \end{tabular}
\end{table*}
\begin{table*}
  \centering
    \caption{All optical observations which were calibrated to the
      $I$-band, showing the central time after trigger of each exposure,
      exposure time ($T_{exp}$), filter,
      instrument, flux and flux errors. The filters quoted are the original
      filters that observations were taken, before conversion to the standard
      filters used in the later analysis. Fluxes are extinction corrected and
      any external references are cited. A full version of this table is
      available online.}
    \label{opt_i}
    \begin{tabular}{@{}ccccccc}
      \hline
      Time after trigger (s) & $T_{exp}$ (s) & Filter & Instrument & Flux ($\mu$Jy) & Flux error ($\pm\mu$Jy) & External sources \\
      \hline
      384 & 60 & \textit{i'} & P60 & 650.01 & 69.31 & \citet{2009ApJ...693.1484C} \\
      642 & 60 & \textit{i'} & P60 & 571.37 & 60.92 & \citet{2009ApJ...693.1484C} \\
      899 & 60 & \textit{i'} & P60 & 587.38 & 33.37 & \citet{2009ApJ...693.1484C} \\
      1246 & 120 & \textit{i'} & P60 & 632.30 & 35.93 & \citet{2009ApJ...693.1484C} \\
      1448 & 60 & \textit{i'} & P60 & 620.76 & 23.30 & \citet{2009ApJ...693.1484C} \\
      1541 & 45 & $I_{c}$ & SMARTS-ANDICAM & 676.00 & 32.00 &
 \citet{2010ApJ...720.1513K} \\
      \hline
    \end{tabular}
\end{table*}
\begin{table*}
  \centering
    \caption{All optical observations which were calibrated to the
      $R$-band, showing the central time after trigger of each exposure,
      exposure time ($T_{exp}$), filter,
      instrument, flux and flux errors. The filters quoted are the original
      filters that observations were taken, before conversion to the standard
      filters used in the later analysis. Fluxes are extinction corrected and
      any external references are cited. A full version of this table is
      available online.}
    \label{opt_r}
    \begin{tabular}{@{}ccccccc}
      \hline
      Time after trigger (s) & $T_{exp}$ (s) & Filter & Instrument & Flux ($\mu$Jy) & Flux error ($\pm\mu$Jy) & External sources \\
      \hline
      57 & 30 & \textit{unfiltered} & KAIT & 29.70 & 9.45 & \\
      162 & 20 & \textit{unfiltered} & KAIT & 571.23 & 10.62 & \\
      171 & 20 & $R$ & Super-LOTIS & 218.11 & 39.33 &
 \citet{2008GCN..7387....1M} \\
      195 & 10 & $R$ & Super-LOTIS & 634.87 & 48.55 &
 \citet{2008GCN..7387....1M} \\
      224 & 36 & $R$ & REM-ROSS & 683.41 & 161.25 & \\
      254 & 20 & \textit{unfiltered} & KAIT & 781.29 & 7.23 & \\
      \hline
    \end{tabular}
\end{table*}
\begin{table*}
  \centering
    \caption{All optical observations which were calibrated to the
      $V$-band, showing the central time after trigger of each exposure,
      exposure time ($T_{exp}$), filter,
      instrument, flux and flux errors. The filters quoted are the original
      filters that observations were taken, before conversion to the standard
      filters used in the later analysis. Fluxes are extinction corrected and
      any external references are cited. A full version of this table is
      available online.}
    \label{opt_v}
    \begin{tabular}{@{}ccccccc}
      \hline
      Time after trigger (s) & $T_{exp}$ (s) & Filter & Instrument & Flux ($\mu$Jy) & Flux error ($\pm\mu$Jy) & External sources \\
      \hline
      84 & 10 & V & UVOT & 246.60 & 140.41 & \\
      113 & 10 & \textit{white} & UVOT & 115.97 & 43.37 & \\
      123 & 10 & \textit{white} & UVOT & 62.36 & 40.39 & \\
      133 & 10 & \textit{white} & UVOT & 102.54 & 43.27 & \\
      143 & 10 & \textit{white} & UVOT & 215.19 & 50.65 & \\
      153 & 10 & \textit{white} & UVOT & 321.34 & 57.17 & \\
      \hline
    \end{tabular}
\end{table*}
\begin{table*}
  \centering
    \caption{All optical observations which were calibrated to the
      $b$-band, showing the central time after trigger of each exposure,
      exposure time ($T_{exp}$), filter,
      instrument, flux and flux errors. The filters quoted are the original
      filters that observations were taken, before conversion to the standard
      filters used in the later analysis. Fluxes are extinction corrected and
      any external references are cited. A full version of this table is
      available online.}
    \label{opt_b}
    \begin{tabular}{@{}ccccccc}
      \hline
      Time after trigger (s) & $T_{exp}$ (s) & Filter & Instrument & Flux ($\mu$Jy) & Flux error ($\pm\mu$Jy) & External sources \\
      \hline
      690 & 10 & $b$ & UVOT & 451.91 & 87.74 & \\
      843 & 10 & $b$ & UVOT & 382.97 & 82.63 & \\
      1465 & 20 & $b$ & UVOT & 406.60 & 79.25 & \\
      5247 & 199 & $b$ & UVOT & 213.69 & 15.71 & \\
      6682 & 199 & $b$ & UVOT & 139.11 & 14.72 & \\
      23481 & 906 & $b$ & UVOT & 40.52 & 4.67 & \\
      \hline
    \end{tabular}
\end{table*}
\begin{table*}
  \centering
    \caption{All optical observations which were calibrated to the
      $u$-band, showing the central time after trigger of each exposure,
      exposure time ($T_{exp}$), filter,
      instrument, flux and flux errors. The filters quoted are the original
      filters that observations were taken, before conversion to the standard
      filters used in the later analysis. Fluxes are extinction corrected and
      any external references are cited. A full version of this table is
      available online.}
    \label{opt_u}
    \begin{tabular}{@{}ccccccc}
      \hline
      Time after trigger (s) & $T_{exp}$ (s) & Filter & Instrument & Flux ($\mu$Jy) & Flux error ($\pm\mu$Jy) & External sources \\
      \hline
      670 & 20 & $u$ & UVOT & 143.74 & 35.38 & \\
      823 & 20 & $u$ & UVOT & 114.74 & 33.43 & \\
      1440 & 20 & $u$ & UVOT & 261.99 & 48.14 & \\
      5041 & 200 & $u$ & UVOT & 152.79 & 11.27 & \\
      6477 & 200 & $u$ & UVOT & 102.75 & 10.28 & \\
      22568 & 906 & $u$ & UVOT & 23.23 & 2.85 & \\
      \hline
    \end{tabular}
\end{table*}
\twocolumn
\label{lastpage}
\end{document}